\documentclass[aps,pre,twocolumn,amsmath,amssymb,notitlepage,nofootinbib,noeprint,superscriptaddress]{revtex4-1}
\pdfoutput=1
\usepackage[latin9]{inputenc}
\usepackage{braket}
\usepackage{xcolor}
\usepackage{verbatim}
\usepackage{graphicx}
\usepackage{esint}
\usepackage{epstopdf}
\usepackage{float}
\usepackage{bm}
\definecolor{darkblue}{rgb}{0,0,0.6}
\usepackage[colorlinks,linkcolor=darkblue,citecolor=darkblue,urlcolor=darkblue]{hyperref}
\newcommand{\red}[1]{\textcolor{black}{#1}}
\newcommand{\beq}{\begin{equation}} \newcommand{\eeq}{\end{equation}}
\newcommand{\argc}[1]{\left[#1\right]}

\begin{document}

\title{Excess wings and asymmetric relaxation spectra in a facilitated trap model}

\author{Camille Scalliet*}

\affiliation{Department of Applied Mathematics and Theoretical Physics, University of Cambridge, Wilberforce Road, Cambridge CB3 0WA, United Kingdom}

\email{cs2057@cam.ac.uk}

\author{Benjamin Guiselin*}

\affiliation{Laboratoire Charles Coulomb (L2C), Universit\'e de Montpellier, CNRS, 34095 Montpellier, France}

\author{Ludovic Berthier}

\affiliation{Laboratoire Charles Coulomb (L2C), Universit\'e de Montpellier, CNRS, 34095 Montpellier, France}

\affiliation{Yusuf Hamied Department of Chemistry, University of Cambridge, Lensfield Road, Cambridge CB2 1EW, United Kingdom}

\let\thefootnote\relax\footnotetext{*Equal contributions.}

\date{\today}

\begin{abstract}
In a recent computer study, we have shown that the combination of spatially heterogeneous dynamics and kinetic facilitation provides a microscopic explanation for the emergence of excess wings in deeply supercooled liquids. Motivated by these findings, we construct a minimal empirical model to describe this physics and introduce dynamic facilitation in the trap model, which was initially developed to capture the thermally-activated dynamics of glassy systems. We fully characterise the relaxation dynamics of this facilitated trap model varying the functional form of energy distributions and the strength of dynamic facilitation, combining numerical results and analytic arguments. Dynamic facilitation generically accelerates the relaxation of the deepest traps, thus making relaxation spectra strongly asymmetric, with an apparent ``excess" signal at high frequencies. For well-chosen values of the parameters, the obtained spectra mimic experimental results for organic liquids displaying an excess wing. Overall, our results identify the minimal physical ingredients needed to describe excess processes in relaxation spectra of supercooled liquids.
\end{abstract}

\maketitle

\section{Introduction}

The large body of experimental work~\cite{ediger1996supercooled,angell1995formation} characterising structural relaxation in supercooled liquids approaching the glass transition reveals that the slow molecular motion is broadly distributed in time~\cite{bohmer1993nonexponential}. Correlation functions (in the time domain) or relaxation spectra (in the frequency domain) demonstrate the existence of molecular dynamics occurring over all timescales separating microscopic motion (in the picosecond range) and the average structural relaxation time $\tau_\alpha$ (about hundreds of seconds at the experimental glass transition temperature $T_g$)~\cite{bohmer1993nonexponential,lunkenheimer2002excess,roland2005supercooled}. A central theme in glass transition studies concerns the physical origin of the rapid growth of $\tau_\alpha(T)$ near $T_g$. However, any physical explanation should also be able to rationalise the main features of the broad distribution of relaxation times characterising the dynamics~\cite{edigerannurev,tarjus2011overview,berthier2011theoretical}.      

Experimentally, various techniques are able to probe the orientational dynamics of molecules over a broad frequency range, such as dielectric spectroscopy~\cite{lunkenheimer2000glassy}, magnetic resonance~\cite{flamig2020nmr}, and dynamic light scattering~\cite{schmidtke2013reorientational}. In most molecular liquids, the structural relaxation is described in the time domain by a stretched (rather than simple) exponential form. In the Fourier domain, this becomes a peak that is broader than a simple Lorentzian, and various empirical functions are used~\cite{bohmer1993nonexponential}. In a large number of liquids, an ``excess" relaxation signal appears in addition to the main $\alpha$-peak on its high-frequency flank~\cite{johari1970viscous,nagel_scaling,nagel_scaling2,menon1992wide,leheny1997high,leheny1998dielectric,adichtchev2003reexamination, blochowicz2003susceptibility, gainaru2009evolution,caporaletti2021experimental,blochowicz2006evolution,gainaru2019spectral}. It is then customary to fit this signal as an additional contribution, or ``process", to the dynamics. Such secondary relaxation may also take different functional forms depending on the studied liquid~\cite{ngai_classification,wu1991relaxation}, but quite often it appears as an extremely broad peak (in the frequency domain), or even a pure power law covering many decades~\cite{nagel_scaling2}. This excess signal then resembles a ``wing" in a log-log representation of the spectra, which are then very asymmetric. The presence of such signal is observed across a wide variety of materials comprising metallic, organic, ionic, polymeric and atomic glasses. The emergence of excess wings is therefore a reasonably universal feature of deeply supercooled liquids.

There are several important questions related to secondary processes in supercooled liquids. Given that the strength of the signal is much smaller than the main $\alpha$-peak (typically 1~\%), \red{two generic explanations may be considered. Either all molecules contribute by performing motions that are about 100 times smaller than during structural relaxation itself~\cite{vogel20022h}, or instead a small fraction of about 1~\% of the molecules undergo structural relaxation much before the bulk.} A third alternative is that secondary relaxation is caused by a slow degree of freedom (say, some intra-molecular motion) that is distinct from the ones contributing to the $\alpha$-peak, which then raises further questions about the inter-relation between these processes. Previous work also tried to identify how universal the shape of the excess signal can be depending on experimental probes~\cite{petzold2013evolution,korber2020systematic}, the type of molecules~\cite{ngai_classification,buchenau2020sound}, and thermodynamic conditions (by, say, varying pressure~\cite{roland2005supercooled}). 

A variety of physical interpretations and empirical models have been proposed to explain the asymmetric shape of relaxation spectra. In the simplest empirical descriptions, the structural and secondary relaxations are described by the linear superposition of two independent contributions~\cite{kudlik1998slow,gainaru2010energy}. This is of course a valid mathematical option, but this does not address the nature of the two relaxation processes, and even implicitly suggests that they can be considered as stemming from independent and distinct molecular motion, which is a questionable hypothesis. 

In most models proposed to describe the glass transition of supercooled liquids, secondary processes do not necessarily appear at the level of the basic formulation~\cite{berthier2011theoretical}. Still, there now exists a good number of proposals regarding the microscopic origin and physical interpretation of excess wings. In a broad, first category of models, asymmetric relaxation spectra emerge when the ``main" glassy degree of freedom responsible for the $\alpha$-relaxation is coupled to an ``additional'' degree of freedom. In Refs.~\cite{diezemann1999slow,mohanty2000nature}, for instance, a trap model describing the structural relaxation is passively coupled to an independent relaxation process, possibly representing a local molecular process. In mode-coupling approaches, the translational degrees of freedom responsible for dynamic arrest are also passively coupled to a second glassy process (for instance mimicking the rotation of molecules) to produce more complex spectra~\cite{gotze1989beta, gotze2002logarithmic, domschke2011glassy, chong2002structural, cummins2005dynamics, sjogren1986diffusion}. In a very similar spirit, the coupling of two schematic glass models was shown to produce spectra with several slow processes~\cite{crisanti2011statistical,crisanti2015simple}, a notable difference with mode-coupling approaches being the more reciprocal interaction between the two glassy processes.

In a second broad family of approaches, asymmetric spectra are directly obtained from an underlying broad distribution of relaxation times with an asymmetric shape. This broad  distribution is meant to describe the heterogeneous nature of glassy relaxation, which is ascribed to static~\cite{sethna1991scaling, stevenson2010universal, viot2000heterogeneous, chamberlin1999mesoscopic, dyre2000universality} or kinetic~\cite{berthier2005numerical} underlying mechanisms. To account for the winged asymmetric shapes of spectra in such approaches, one needs to invoke at least one physical ingredient. In the approach based on geometric frustration~\cite{tarjus2005frustration}, large correlated domains are more energetically suppressed than small ones due to frustration while they all relax via thermal activation, which accounts for excess wings~\cite{viot2000heterogeneous}. In the model developed by Chamberlin~\cite{chamberlin1993non,chamberlin1998experiments,chamberlin1999mesoscopic} the distribution of domain sizes is symmetric, but large and small domains relax differently~\cite{chamberlin1999mesoscopic}, thus resulting in asymmetric time distributions. In some dynamically facilitated models~\cite{garrahan2011kinetically,chandler2010dynamics}, the structural relaxation emerges as a hierarchical process involving a distribution of timescales which is skewed towards small times, giving rise to asymmetric spectra in the Fourier domain~\cite{berthier2005numerical}. In the mosaic (droplet) picture of the random first order transition theory~\cite{kirkpatrick1989scaling,lubchenko2007theory}, secondary processes result from fluctuations of the droplet sizes and shapes~\cite{stevenson2010universal}. They thus induce an asymmetric distribution of free energy barriers overcome by thermal activation. We note that the impact of dynamic facilitation on the dynamics of the largest droplets was also invoked in this context~\cite{xia2001microscopic, bhattacharyya2008facilitation,biroli2012random,berthier2019can}, but this was not related to the existence of excess wings.   

In a recent computational study~\cite{shortwings}, we investigated the equilibrium relaxation dynamics of a simple supercooled liquid thermalised down to the experimental glass transition temperature $T_g$ using the swap Monte Carlo algorithm~\cite{berthier2016equilibrium,swap,berthier2019efficient}, and we were able to follow the relaxation dynamics of the particles over a time window of about 10 orders of magnitude. In particular, we could access the temperature regime where excess wings are observed experimentally and record the dynamics over the relevant time window as well. We demonstrated the emergence of an excess wing in this deeply supercooled regime, and observed the microscopic particle motion responsible for the asymmetric shape of the relaxation spectra. Our main conclusion was that relaxation close to $T_g$ is the result of two central physical ingredients, dynamic heterogeneity and dynamic facilitation. First, relaxation is initiated over timescales much shorter than $\tau_\alpha$ at a sparse population of localised regions, these relaxation events being extremely broadly distributed. The dynamics is therefore spatially and temporally heterogeneous in this regime. These sparse relaxed regions were found to trigger the relaxation of neighbouring regions by kinetic facilitation.  We argued that since dynamic facilitation accelerates the relaxation of the slowest regions in the liquid, the broad relaxation spectrum observed at high frequencies becomes compressed at low frequencies, thus explaining the asymmetric shape of the observed spectra. 

In our previous work~\cite{shortwings}, we suggested that a simple facilitated trap model combining these two ingredients (heterogeneity and facilitation) should generically lead to asymmetric relaxation spectra. Here, we provide a complete study of the relaxation dynamics of the proposed model. We describe how this facilitated trap model is actually constructed, by making the simplest possible assumptions. We then extensively explore the parameter space. We show that, despite its extreme simplicity, the model naturally and generically gives rise to asymmetric, winged relaxation spectra in the presence of dynamic facilitation. We also propose a microscopic analysis of the model itself, combining detailed numerical simulations to an approximate analytic solution. Combined to our numerical simulations~\cite{shortwings}, these results provide a simple, physically-motivated microscopic picture for the shape of relaxation spectra in deeply supercooled liquids.

Our article is organised as follows. In Sec.~\ref{sec:Model}, we introduce the facilitated trap model and dynamic observables. In Sec.~\ref{sec:spectra}, we obtain the relaxation spectra. We present in Sec.~\ref{sec:microscopicfacilitation} our analysis of the individual trap dynamics to rationalise the spectral shapes. In Sec.~\ref{sec:wings} we summarise our results and discuss the microscopic picture that explains the asymmetric, winged relaxation spectra of deeply supercooled liquids which emerges from our study, in relation to both experiments and previous models. 

\section{Facilitated trap model}

In this section we define the facilitated trap model, along with observables employed to investigate its dynamics. We also describe our numerical simulations of the model.

\label{sec:Model}

\subsection{Original trap model}

\begin{figure}[t]
\centering{
\includegraphics[width=\columnwidth]{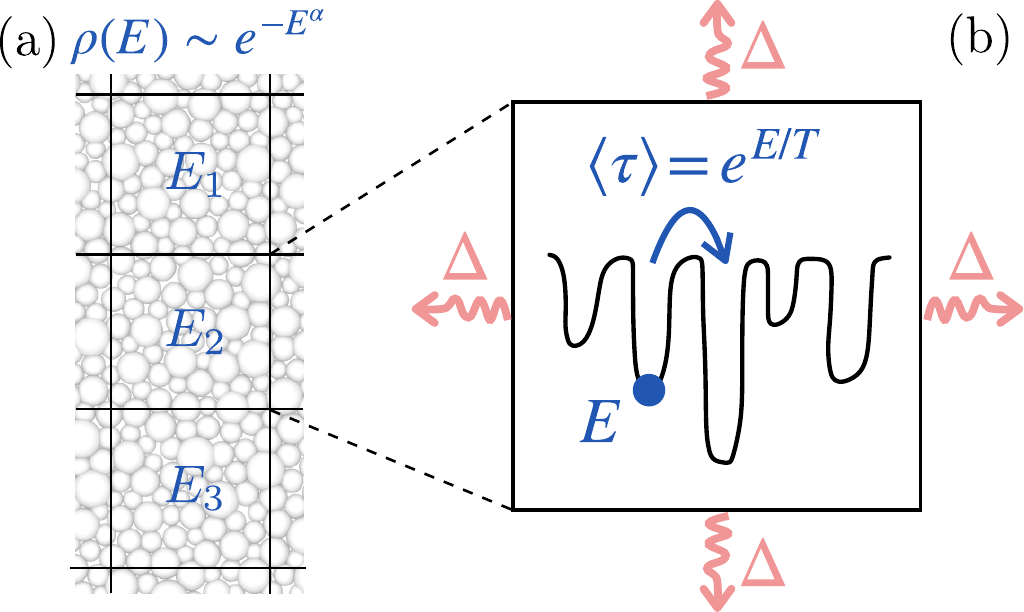}}
\caption{Facilitated trap model. (a) The liquid is coarse-grained as a collection of traps of depth $E$ distributed according to $\rho(E)$. Hopping of a mesoscopic domain is a thermally-activated process. (b) Whenever a trap relaxes, it perturbs the energy of all other traps by an amount proportional to $\Delta$.} 
\label{fig:model}
\end{figure}

The dynamics of deeply supercooled liquids exhibits very strong dynamic heterogeneities~\cite{edigerannurev, berthier2011dynamical,berthier2011physics}. This suggests that the supercooled liquid can be coarse-grained into independently relaxing domains characterised by a local relaxation time that is broadly distributed, as illustrated in Fig.~\ref{fig:model}(a). This spatially heterogeneous viewpoint is mathematically captured by the original trap model~\cite{brawer1984theory,dyre1987master,bassler1987viscous,bouchaud1992weak,Monthus_1996,comtetmonthus,denny2003trap,heuer2005potential,diezemann2011memory}. Our computer simulations of a simple atomistic glass-forming model has revealed the heterogeneous nature of the particle motion that gives rise to the high-frequency part of the relaxation spectra~\cite{shortwings}. The trap model is therefore a natural starting point. 

The original trap model describes the thermally-activated dynamics of $N$ mesoscopic domains hopping between energy traps of depth $E\geq 0$, with or without any spatial structure. \red{This description makes it clear that a direct connection to detailed molecular motion in the original system is difficult.} At a temperature $T$, the domains escape the trap they occupy after a Poisson-distributed relaxation time $\tau$ of mean $\tau_0 e^{ \beta E}$, with $\beta=1/T$ the inverse temperature (the Boltzmann constant is set to unity) and $\tau_0$ a microscopic time constant, as illustrated in Fig.~\ref{fig:model}(b). Dynamic heterogeneity is introduced through the distribution $\rho(E)$ of trap depths. We consider a general functional form for this distribution
\beq 
\rho(E) = \frac{\alpha}{\Gamma(1/\alpha)E_0}e^{-(E/E_0)^{\alpha}},
\label{eq:rho}
\eeq
with $E_0$ a constant energy scale, and an exponent $1 < \alpha \leq 2$ to smoothly interpolate between the exponential~\cite{bouchaud1992weak,Monthus_1996,comtetmonthus} and Gaussian~\cite{brawer1984theory,dyre1987master,bassler1987viscous,diezemann1997free, rehwald2010coupled} distributions. Both these models have been widely studied in various contexts. B\"assler and coworkers~\cite{arkhipov1994random,arkhipov1996random} have previously introduced the idea of an exponent $\alpha$ which varies continuously, in order to perform quantitative comparisons to some experimental results. 

When a given domain relaxes and jumps out of its trap, it chooses a new trap depth $E'$ with probability $\rho(E')$ without memory of the previous energy $E$. The resulting dynamics of the model thus depends on two parameters: $\alpha$, which specifies the distribution $\rho(E)$, and the temperature $T$. We express energies and times in units of $E_0$ and $\tau_0$, respectively, and consider a mean-field version with no spatial structure or interaction between traps.  

In thermal equilibrium, the probability distribution $P_{\text{eq}}(E)$ for a sub-system to be in a trap of depth $E$ at the temperature $T$ is given by~\cite{Monthus_1996}
\beq 
P_{\text{eq}}(E) = \frac{\rho(E) e^{ \beta E}}{Z(\beta) },
\eeq
with $Z(\beta)  = \int_0^{\infty} \rho(E) e^{ \beta E} \mathrm{d} E$ a normalisation factor which is finite and strictly positive at all $T>0$ as long as $\alpha > 1$ (the exponential trap model with $\alpha=1$ has instead a finite temperature phase transition to a non-ergodic low-temperature phase~\cite{bouchaud1992weak}). We define the average trap depth $\overline{E}(T)$ via $\overline{E} = \int_0^{\infty} P_{\text{eq}}(E) E \mathrm d E$.

By considering the thermally-activated relaxation over distributed energy barriers, the trap model naturally gives rise to dynamic heterogeneities. Sub-systems in shallow traps $E \ll \overline{E}$ relax much faster than the ones stuck in deep traps with $E \gg \overline{E}$. The parameter $\alpha$ directly controls the breadth of dynamic heterogeneities: smaller $\alpha$ corresponds to broader distribution $\rho(E)$, and thus to much wider distribution of local relaxation times.

\subsection{Facilitated trap model}

The above model considers that sub-systems are dynamically independent, and that the motion out of a given trap does not affect the state of the other domains. This is not true in realistic glass-formers, in which a relaxation event corresponds to a local rearrangement of particles, which inevitably modifies the environment of the neighbouring particles~\cite{chandler2010dynamics}. Computer simulations reveal that this effect facilitates the gradual spreading of relaxation events through the liquid~\cite{bergroth2005examination,vogel2004spatially,candelier2009building,candelier2010spatiotemporal,keys2011excitations,schoenholz2016structural}, and plays an increasingly important role at low temperatures~\cite{chacko} (an opposite conclusion was reached in Ref.~\cite{candelier2009building}). In our numerical investigation of excess wings, we also found that the initial relaxation in isolated locations facilitates the relaxation of nearby regions~\cite{shortwings}. 

While most studies of the trap model ignore the effect of dynamic facilitation, we note that the initial studies of the exponential trap model in Refs.~\cite{comtetmonthus, Monthus_1996} had actually introduced such dynamic correlation between traps. A Gaussian trap model with dynamic exchanges between fast and slow regions was studied in Ref.~\cite{diezemann2005aging} to analyse the consequences of spatially heterogeneous dynamics in supercooled liquids. More recently, Heuer and coworkers have also included a similar effect in a lattice version of the model, in order to describe the finite-size dependence of the dynamics of supercooled liquids~\cite{rehwald2010coupled, rehwald2012coupled}. Here we follow a similar philosophy and also assume that every hopping event induces a small shift in the depth of the neighbouring traps, as illustrated in Fig.~\ref{fig:model}(b). 

To simplify the analytic description of this facilitation effect we assume, again in a mean-field spirit, that all traps are equally affected by hopping processes. In practice, whenever a trap relaxes, we attempt to shift the energy of all other traps by a random amount $\delta E$, different for each trap, uniformly distributed in the interval $\argc{- \frac{\Delta}{2\sqrt{N}}, \frac{\Delta}{2\sqrt{N}}}$: $E \rightarrow E'=E + \delta E$. The energy scale $\Delta$ controls the strength of dynamic facilitation ($\Delta=0$ gives back the original trap model) and the scaling with $N$ ensures that the resulting dynamics scales correctly with the system size. The change in energy is then accepted or rejected in order to leave the equilibrium probability distribution $P_{\text{eq}}(E)$ unchanged. This facilitation effect thus corresponds to a drift-diffusion process of the trap depth in energy space in an effective confining potential (in appropriate units) $V_{\rm eff}= -\log P_\mathrm{eq}$ (see Sec.~\ref{subsec:approxdynamics} for an analytic description). If the new trap energy $E'$ is accepted, we assign a new Poisson-distributed relaxation time of mean $e^{\beta E'}$ to that site. At fixed $\alpha$ and $T$, increasing $\Delta$ corresponds to increasing the strength of dynamic facilitation. \red{Notice that due to dynamic facilitation, the energy of a given trap can both increase or decrease in a stochastic manner. However, since the energy diffusion occurs in a potential $V_{\rm eff}$, the energy of traps with a large initial depth will slowly return back to the average energy value. We shall show below that such dynamic facilitation leads to an average speedup of the global dynamics, which is not introduced by hand in the formulation of the model.}

\subsection{Dynamic observables}

By construction, the equilibrium distribution of energies  $P_{\text{eq}}(E)$ does not depend on the presence of dynamic facilitation and is independent of $\Delta$. Although thermodynamic properties do not depend on $\Delta$ we expect the dynamic relaxation to be strongly affected by the presence of dynamic facilitation. 

To investigate this effect, we define a time correlation function to describe the dynamics of the traps. To this end, we introduce the mean persistence $p(t)$, which quantifies the fraction of sub-systems which, at time $t$, have not escaped the trap they occupied at time 0. The persistence function monotonically decreases from one at $t=0$ to zero at long times, when all sub-systems have relaxed. It is physically equivalent to the self-intermediate scattering function at large wavevectors, which is more typically used in off-lattice glass models. The mean persistence $p(t)$ can be expressed as an average of the persistence $p(t;E_i)$ of a single sub-system starting from an initial trap depth $E_i$ at $t=0$ over the equilibrium distribution $P_\mathrm{eq}(E_i)$,
\beq
p(t) = \int_{0}^{\infty} \mathrm  P_{\text{eq}}(E_i) p(t;E_i) \mathrm{d}E_i.
\label{eq:persistence}
\eeq
In the absence of dynamic facilitation, the persistence of a single sub-system can be derived exactly from the exponential distribution of relaxation times and reads $p(t;E_i) = e^{-t e^{-\beta E_i}}$. The global persistence can thus be computed analytically from Eq.~(\ref{eq:persistence}). In the presence of dynamic facilitation $\Delta >0$, the persistence of a single sub-system $p(t;E_i)$ stems from a non-trivial interplay between the activated jump dynamics and the facilitated, diffusive, dynamics of the trap depth in energy space, and cannot be computed analytically exactly. 
In Sec.~\ref{subsec:approxdynamics} we provide an approximate analytic solution for $p(t)$ in the presence of facilitation. The exact persistence of the facilitated trap model is thus measured by means of numerical simulations using 
\beq
p(t) = \left \langle \frac{1}{N} \sum_{i=1}^N p_i(t) \right\rangle,
\label{eq:persistence_simu}
\eeq
where $p_i(t) =1$ if the sub-system $i$ has not escaped at time $t$ from the trap it occupied at time 0, and $p_i(t)=0$ otherwise. The brackets indicate an average over independent initial conditions (trap energies) drawn from the equilibrium distribution $P_{\rm eq}(E)$.

In experiments, glassy features are often investigated thanks to spectroscopic techniques. In order to allow for a direct comparison, we introduce a dynamic susceptibility $\chi''(\omega)$ in the frequency domain. Following the experimental literature, we assume that the dynamics can be described by a distribution of relaxation times $G(\log \tau)$ such that~\cite{bohmer1998nature, blochowicz2003susceptibility, berthier2005numerical}
\begin{equation}
\chi''(\omega)=\int_{-\infty}^{\infty} G(\log \tau) \frac{\omega \tau }{1+(\omega \tau)^2}\mathrm{d}\log \tau,
\label{eq:chiG}
\end{equation} 
with $\omega$ the frequency (up to a factor $2\pi$), and where the double prime indicates that we consider the imaginary part of the total susceptibility. The global persistence corresponds to the superposition of step functions decaying from 1 to 0 when a sub-system jumps. As a consequence, the relation $G=-\mathrm{d}p/\mathrm{d}\log t$ holds~\cite{berthier2005numerical} and the susceptibility spectrum $\chi''(\omega)$ can be directly computed from the persistence data:
\begin{equation}
\chi''(\omega) =  - \int_{-\infty}^{+\infty}\frac{\mathrm{d}p }{\mathrm{d}\log t}(\log \tau) \frac{\omega\tau}{1+(\omega \tau)^2}\mathrm{d}\log \tau,
\label{eq:chip}
\end{equation}
for any value $\Delta \geq 0$. In the case $\Delta=0$, we directly evaluate the persistence given by Eq.~(\ref{eq:persistence}). For $\Delta>0$, we instead use the persistence measured in the simulations via Eq.~(\ref{eq:persistence_simu}). Both are then used to compute $\chi''(\omega)$ via Eq.~(\ref{eq:chip}).

\subsection{Simulations of the facilitated trap model}

When $\Delta>0$, the persistence cannot be computed analytically exactly from Eq.~(\ref{eq:persistence}) and we instead use simulations and Eq.~(\ref{eq:persistence_simu}) to measure the time correlation function. We provide the details of these simulations. 

First we initialize the $N$ trap energies by directly sampling the equilibrium distribution $P_{\text{eq}}(E)$. To do so, we numerically evaluate the cumulative probability distribution of energies $\mathcal{C}_{\text{eq}}(E)=\int_0^EP_\mathrm{eq}(E')\mathrm{d}E'$ on a grid of a thousand points from $E_\mathrm{min}$ where $\mathcal{C}_\mathrm{eq}(E_\mathrm{min})=10^{-10}$, to $E_\mathrm{max}$ where $\mathcal{C}_\mathrm{eq}(E_\mathrm{max})=1-10^{-10}$, as ${\cal C}_{\rm eq}$  cannot be computed analytically for arbitrary values of $\alpha$. We use a cubic spline interpolation to construct numerically the reciprocal function $\mathcal{E}={{\mathcal C}_{\rm eq}}^{-1}$. For each trap, we generate a random variable $X$ from a uniform distribution in the range $\argc{0,1}$, and we assign it an initial energy $E_i = \mathcal{E}(X)$. 

The relaxation time of a trap of depth $E$ is drawn from an exponential distribution of mean $e^{\beta E}$. To accomplish this, we generate a random number $X$ from a flat distribution in the range $\argc{0,1}$ and construct the relaxation time $\tau = - e^{\beta E}\log(X)$. 

In the course of the simulation, when a trap relaxes, we sample the probability distribution $\rho(E)$ defined in Eq.~(\ref{eq:rho}) to select its new energy. For this purpose, we generate two random variables: $X$ from a flat distribution  in the range $\argc{0,1}$ and $Y$ from the Gamma distribution $\Gamma(1+1/\alpha,1)$ with shape parameter 1 + $1/\alpha$ and rate parameter 1. The resulting energy $E=X Y^{1/\alpha}$ is then distributed according to $\rho(E)$~\cite{gamma}.

Facilitation in the model corresponds to a diffusion process in energy space that preserves the equilibrium distribution of trap energies, hence which is confined to an effective potential $V_{\rm eff} = -\log P_{\text{eq}}$ (in appropriate units). To this end, we compute the difference in effective potential that a proposed shift of energy would imply, $\delta V_{\rm eff} = \log (P_{\rm eq}(E) / P_{\rm eq}(E'))$ and use a Metropolis filter. If $\delta V_{\rm eff} < 0$, we accept the new trap energy $E'$. If $\delta V_{\rm eff} > 0$, the change is accepted with probability $e^{-\delta V_{\rm eff}}$.

To measure the global persistence given by Eq.~(\ref{eq:persistence_simu}), we perform numerical simulations for $N=100$ until full decorrelation is reached, using $\Delta$ values in the range $\argc{10^{-4},1}$. Since the persistence decreases by steps of size $1/N$, large systems of size $N=10^4$ are used to resolve the persistence function at very short times, when needed. We then combine the persistence data measured at short times in very large systems with the data measured in smaller systems at longer times. We typically average the persistence over 100 independent simulations for $N =100$, and 10 runs for $N=10^4$ in order to obtain sufficient statistics.
 
\section{Relaxation spectra}
\label{sec:spectra}

In this section, we first report the relaxation spectra of the trap model in the absence of dynamic facilitation for $\Delta=0$. We present results for various underlying distributions of energy barriers $\rho(E)$, parametrized by $\alpha$, and for different temperatures $T$. We then present the effect of dynamic facilitation, characterised by $\Delta$, on the relaxation spectra.

\subsection{Original trap model without facilitation}

\label{subsec:spectranofacilitation}

\begin{figure}[t]
\includegraphics[width=\columnwidth]{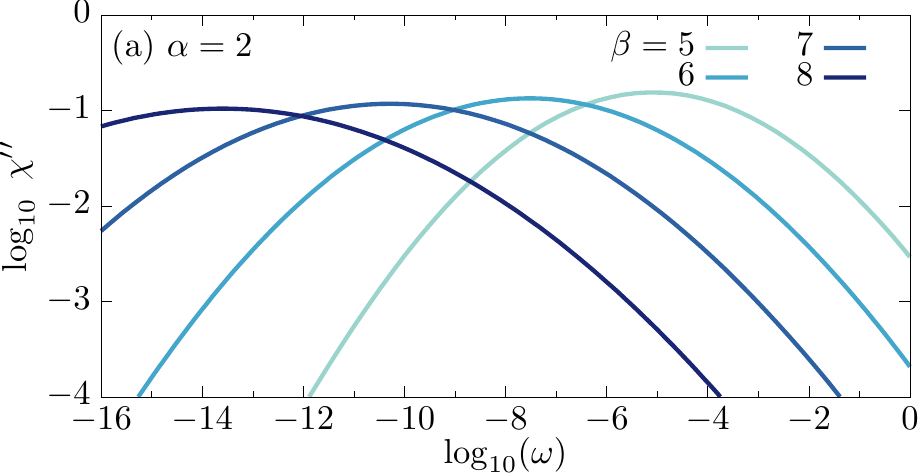}
\includegraphics[width=\columnwidth]{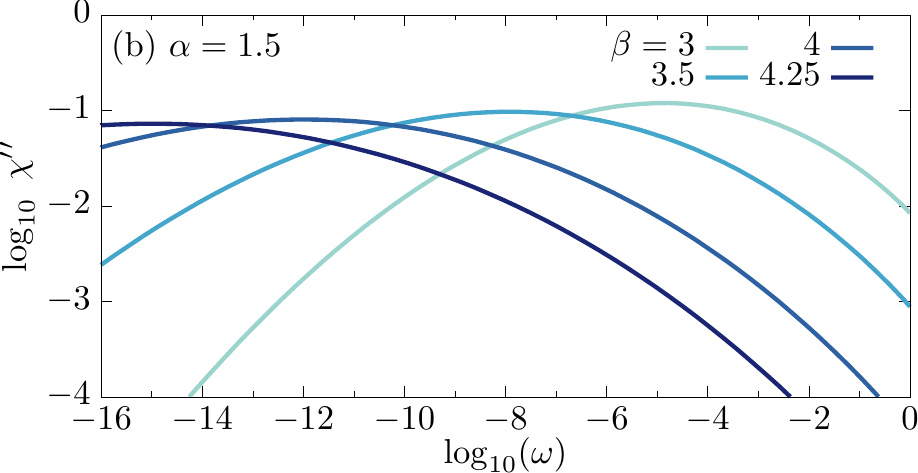}
\includegraphics[width=\columnwidth]{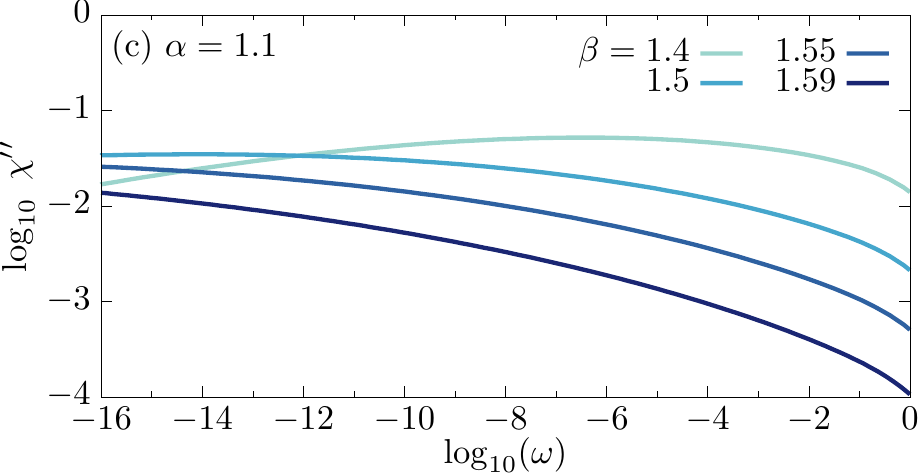}
\caption{Relaxation spectra of the original trap model in the absence of facilitation for different distributions of trap energies $\rho(E) \sim e^{-E^\alpha}$: (a) $\alpha = 2$, (b) $\alpha=1.5$ and (c) $\alpha=1.1$, and various temperatures $T=1/\beta$. For a given value of $\alpha$, the spectra broaden and shift to lower frequencies with decreasing temperature. In addition, the spectra also broaden moving from a Gaussian (a) to an almost-exponential (c) distribution $\rho(E)$ of trap energies.}
\label{fig:nofacilitation} 
\end{figure}

We consider the relaxation spectra of the trap model in the absence of dynamic facilitation ($\Delta=0$) to illustrate the variety of spectral shapes exhibited by the model.

We present in Fig.~\ref{fig:nofacilitation} the relaxation spectra obtained for $\alpha = 2$, 1.5, and  1.1 at various temperatures. The former corresponds to the Gaussian model, which has been studied in the past, while the latter is very close to an exponential distribution for $\rho(E)$. In order to highlight quantitative differences, we show the data over the same frequency and susceptibility ranges in all cases. We also restrict the frequencies to an experimentally-relevant regime, extending over about 14 decades. We observe that all spectra, at various $\alpha$ and temperatures $T$, share a qualitatively similar shape: they are broad, spanning many orders of magnitude in the frequency domain, and are relatively symmetric around their maximum value. The broadness of the spectra is a direct signature of dynamic heterogeneity, and reflects the fact that the sub-systems relax over a broad distribution of timescales.

Yet, we notice from Fig.~\ref{fig:nofacilitation} a clear quantitative evolution of the spectra as a function of $\alpha$ and $\beta=1/T$. At fixed $\alpha$, the spectra shift to lower frequencies and broaden as the temperature decreases. This is due to thermal activation, which gives rise to longer relaxation times and simultaneously enhances dynamic heterogeneity at low temperatures. The exponent $\alpha$ directly controls how broad the underlying distribution of trap energies $\rho(E)$ is, and therefore how strong dynamic heterogeneities are, irrespective of the temperature. While a Gaussian distribution $\rho(E)$ for $\alpha=2$ yields relatively narrow spectra, see Fig.~\ref{fig:nofacilitation}(a), the almost-exponential distribution $\rho(E)$ for $\alpha=1.1$ generates the extremely broad spectra shown in Fig.~\ref{fig:nofacilitation}(c), which extend to physically unreachable low frequencies at low temperatures, for example for $\beta=1.59$, where the entire spectrum does not fit the experimental frequency window.  

These results confirm the salient and well-known features of the trap model that generically leads to slow dynamics and broad relaxation spectra, which appear, however relatively featureless \red{with a main relaxation peak in the Fourier domain. Notice that the spectra found within the trap model are very broad (the behaviour at low frequencies is typically slower than a power law and depends on both $\alpha$ and $T$) and temperature dependent (so that time temperature superposition is not obeyed).}

\subsection{Effect of dynamic facilitation}

\label{subsec:spectrafacilitation}

We now investigate how the introduction of facilitation, parametrized by $\Delta>0$, affects the spectral shapes. We have systematically investigated the effect of $\Delta$ on the dynamics for $\alpha \in ] 1, 2 ]$ and various temperatures. We present results for well-chosen values of $\alpha$, $\beta$, and varying $\Delta$, to illustrate the variety of observed behaviours. 

\begin{figure}[t]
\includegraphics[width=0.48\textwidth]{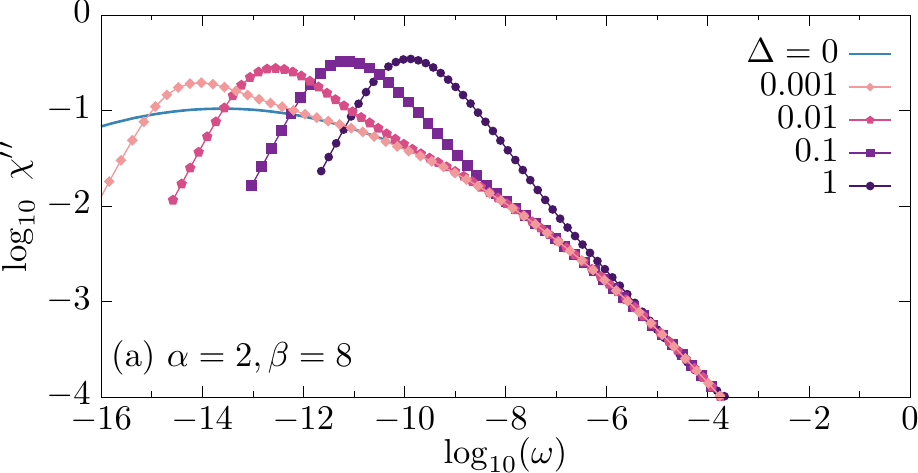}
\includegraphics[width=0.48\textwidth]{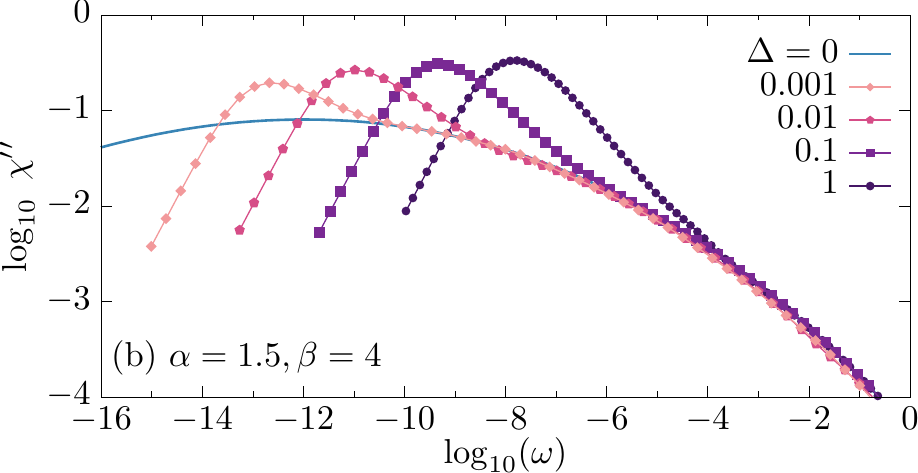}
\includegraphics[width=0.48\textwidth]{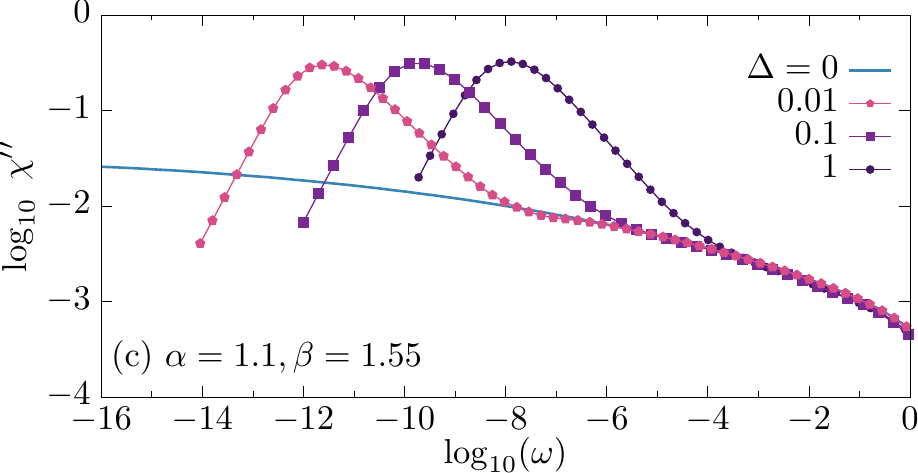}
\includegraphics[width=0.48\textwidth]{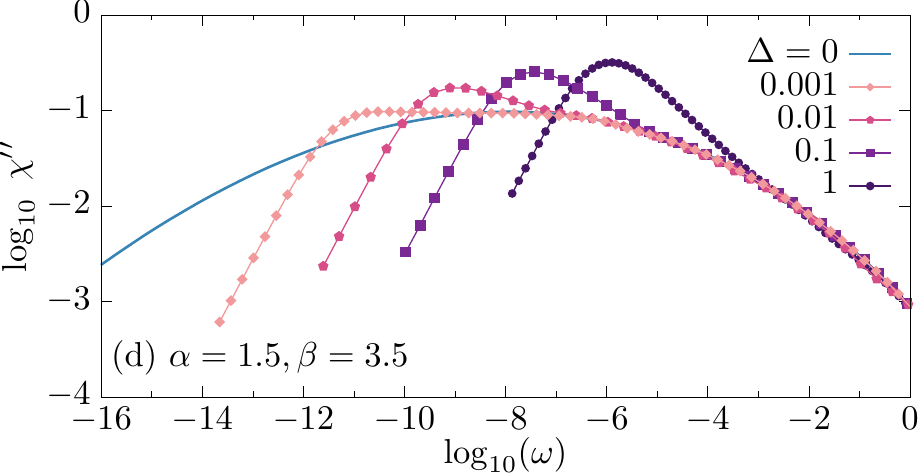}
\caption{Effect of dynamic facilitation, parametrized by $\Delta$, on the relaxation spectra with parameters: (a) $\alpha = 2$, $\beta = 8$, (b) $\alpha = 1.5$, $\beta = 4$, (c) $\alpha = 1.1$, $\beta = 1.55$, (d) $\alpha = 1.5$, $\beta = 3.5$. }
\label{fig:effectdelta} 
\end{figure}

We present in Fig.~\ref{fig:effectdelta} the relaxation spectra of the facilitated trap model for $\alpha =2$, $1.5$, and $1.1$. For each $\alpha$, we choose a relatively low temperature, for which the spectrum without facilitation is very broad and thus for which facilitation is expected to have a larger impact. In Fig.~\ref{fig:effectdelta}(a), we show the spectra for the Gaussian model ($\alpha =2$) at an inverse temperature $\beta = 8$. The spectrum without facilitation $\Delta = 0$ (solid line) is broad, relatively symmetric, and reaches a maximum at $\log_{10} \omega_\alpha \sim -14$. We show the data for $\Delta$ increasing logarithmically from $0.001$ to $1$ (left to right). At the lowest $\Delta=0.001$, the data coincides with $\Delta = 0$ at high frequencies, but deviates from it below $\log_{10} \omega \sim -12$. We see that low frequencies are suppressed by $\Delta >0$, causing a compression of the spectrum into a peak around $\log_{10} \omega_p \sim -14$. The resulting peak is much sharper than the broad hump of the underlying $\Delta=0$ curve. As $\Delta$ increases, the low-frequencies are increasingly suppressed, and the resulting peak increasingly sharper, shifting to a higher $\omega_p$. The frequency range over which the data coincide with $\Delta = 0$ decreases with increasing $\Delta$.

The suppression of low frequencies and the emergence of a sharper peak is a generic effect of imposing $\Delta >0$ in the facilitated trap model. This is confirmed by the results in Figs.~\ref{fig:effectdelta}(b,c) obtained with $\alpha=1.5$, $\beta=4$ and $\alpha=1.1$, $\beta=1.55$, respectively. They show a similar trend to the Gaussian case in Fig.~\ref{fig:effectdelta}(a). In the three first panels, facilitation gives rise to a sharp peak at $\omega_p > \omega_\alpha$ (the latter being the location of the maximum in the $\Delta=0$ spectrum). This peak shifts to higher frequencies as the strength of facilitation is raised. On the other hand, spectra follow the underlying $\Delta = 0$ signal at high frequencies, resulting in a highly asymmetric spectrum, in great contrast with the symmetric spectra without facilitation shown in Fig.~\ref{fig:nofacilitation}.
Finally, we note that the effect of facilitation is more spectacular at small $\alpha$, where the $\Delta=0$ spectrum is extremely broad, extending to unphysically small frequencies. The relative amplitude of the peak is almost two orders of magnitude larger than the underlying signal at $\Delta = 0$ in this case, while it represents about one order of magnitude for $\alpha=1.5$, and even less for $\alpha=2$. 

Finally, we show in Fig.~\ref{fig:effectdelta}(d) the spectra obtained for $\alpha = 1.5$, $\beta = 3.5$. The exponent $\alpha$ is the same as in panel (b), but the temperature is higher, explaining why the $\Delta = 0$ spectrum is centered around a larger frequency $\log_{10}\omega_\alpha \sim -8$. Here, a small $\Delta = 0.001$ also compresses the low-frequency part of the spectrum, giving rise to a secondary peak but the effect is however extremely weak. As $\Delta$ increases, the peak clearly emerges, and shifts to higher frequencies as for the lower temperature. The comparison between panels (b) and (d) reveals that for a given value of $\Delta$, the effect of facilitation is more pronounced at lower temperature and depends on both $\Delta$ and $T$. This point is further addressed in Sec.~\ref{subsec:approxdynamics}.

Overall, this analysis demonstrates that dynamic facilitation compresses the low-frequency part of the relaxation spectra, but leaves the high-frequency regime unaffected. This generically gives rise to highly asymmetric shapes composed of a relatively sharp peak at a low frequency $\omega_p$ mainly controlled by $\Delta$, and a broader shape on the high-frequency flank, controlled by the underlying distribution $P_{\rm eq}(E)$, and thus by the exponent $\alpha$. Whereas the high-frequency regime of the spectra shown in Fig.~\ref{fig:effectdelta} \red{depends on the control parameters, the low-frequency regime is identical with a linear frequency dependence, expected when the underlying distribution of relaxation times is bounded at large times. The linear frequency dependence is not obeyed in the original trap model for $\Delta=0$. We have not systematically investigated whether time temperature superposition holds for the facilitated trap model, but we recall that some models with kinetic facilitation do obey superposition while some others do not, the difference lying in the details of the kinetic facilitation rules~\cite{garrahan2011kinetically}.}    

\section{Microscopic analysis of the facilitated trap model}

\label{sec:microscopicfacilitation}

In this section we investigate the dynamics of individual traps in the presence of dynamic facilitation to rationalise the evolution of the relaxation spectra characterised by an emerging peak frequency $\omega_p$ and an asymmetric winged shape. 

\subsection{Origin of asymmetric spectra}

\label{subsec:physicaloriginfacilitation}

\begin{figure}[t]
\centering{
\includegraphics[width=\columnwidth]{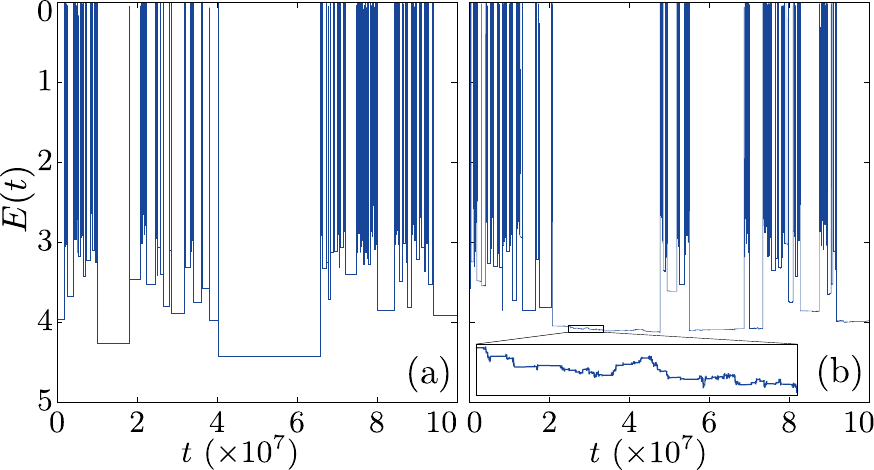}
\includegraphics[width=\columnwidth]{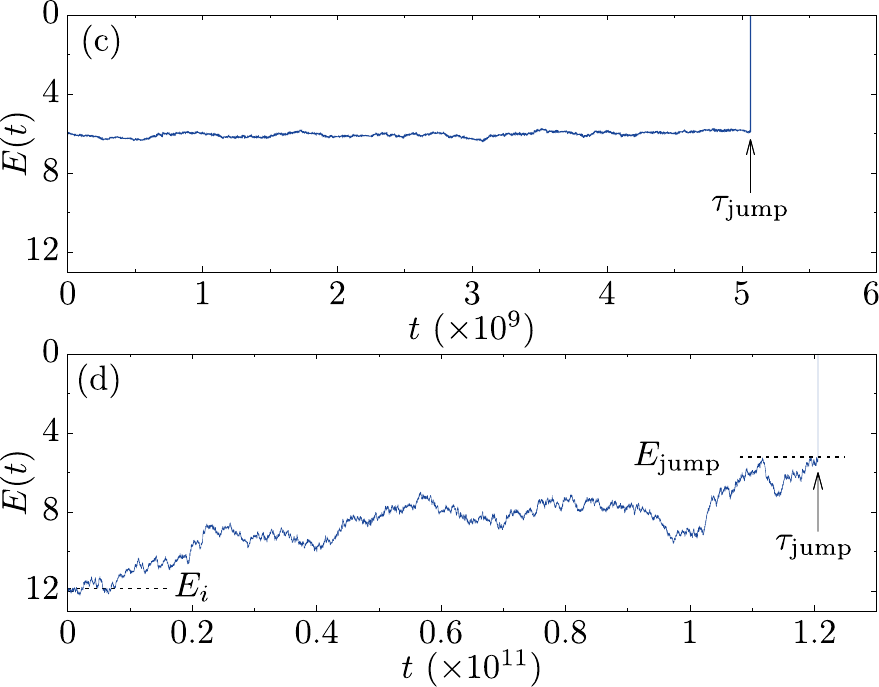}}
\caption{Representative time series of a given trap depth $E(t)$ in the absence (a) and presence \red{(b-d)} of dynamic facilitation. The vertical axis is reversed so that large $E$ values appear as ``deep'' traps. While the time series (a, b) look similar, the inset in (b) reveals the slow diffusion of the energy in deep traps. \red{(c,d) Time evolution of a trap starting with an energy $E_i$ at $t=0$ in the presence of facilitation. The effect of $\Delta >0$ for $E_i = 6$ is little (c), but accelerates greatly the dynamics of a deep trap with $E_i=12$ (d), which undergoes its next jump when its energy is $E_{\rm jump}$ at a time $\tau_{\rm jump} \ll e^{\beta E_i}$}. Data are obtained for $\alpha = 1.5$, $\beta = 4$, $\Delta = 0.01$.} 
\label{fig:series}
\end{figure} 

We illustrate the dynamics of a single trap with and without dynamic facilitation in Fig.~\ref{fig:series}, for $\alpha = 1.5$, $\beta = 4$. In the original trap model, see Fig.~\ref{fig:series}(a), the time series is a succession of plateaus corresponding to the residence times at each given depth. As expected, shallow traps are short-lived, while deep ones are much longer-lived. 

When $\Delta >0$, the trap energy constantly receives energy kicks when other traps hop. The time series is therefore composed of sharp jumps when the energy of the probed trap is renewed, but its energy now slowly evolves in between the jumps. The inset in Fig.~\ref{fig:series}(b) magnifies these small fluctuations. The effect of a small $\Delta > 0$ on the dynamics of shallow traps is negligible. 

We display in Fig.~\ref{fig:series}(c) the evolution of the energy for a trap starting from a large initial depth $E_i$ for the same parameters. In particular the value $E_i\sim 12$ is larger than the mean energy value $\overline{E}\sim 7.2$. For $\Delta=0$, the energy would remain constant over a typical time $e^{\beta E_i} \sim 10^{20}$. When $\Delta = 0.01$, the energy slowly drifts towards lower values, closer to the mean. From this time evolution, we define the energy $E_{\text{jump}}$ of the trap when it jumps, along with the time $\tau_{\text{jump}}$ before it jumps, see Fig.~\ref{fig:series}(c). Remarkably, we have that $\tau_{\text{jump}} \ll e^{\beta E_i}$. Therefore, adding facilitation dramatically accelerates the relaxation of deep traps.
In this specific example, we also note that $\tau_{\text{jump}} \neq e^{\beta E_{\text{jump}}}$. In general we find that for large $E_i$, we have $\tau_{\text{jump}} \gg e^{\beta E_{\text{jump}}}(\sim 10^8)$. This suggests that the relaxation mechanism of the deepest traps is ruled by the diffusion of the trap depth  in energy space: the trap energy slowly diffuses until it explores low values which allow for its relaxation. 

\begin{figure}[t]
\includegraphics[width=0.494\columnwidth]{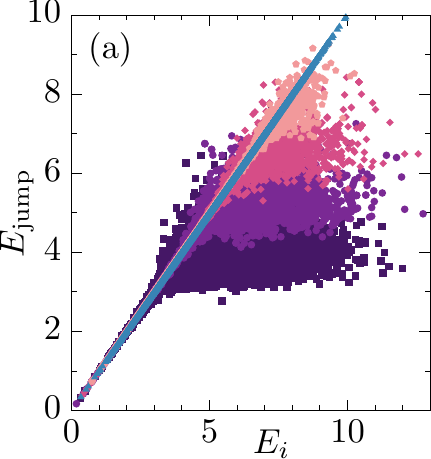}
\includegraphics[width=0.494\columnwidth]{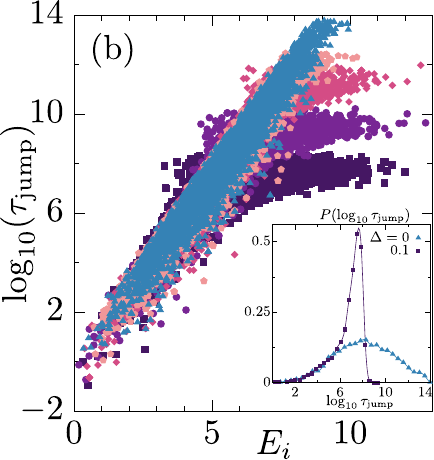}
\caption{Scatter plot of (a) the energy $E_{\text{jump}}$ and (b) time $\tau_{\text{jump}}$ at which traps of initial energy $E_i$ jump. Data are for $\alpha=1.5$, $\beta = 3.5$, and from top to bottom $\Delta = 0,\ 10^{-4},\ 10^{-3},\ 10^{-2},\ 10^{-1}$. The traps relax at smaller energies and timescales with increasing $\Delta$, on average. \red{The inset in (b) collects the data for $\tau_{\rm jump}$ in the distribution $P(\log_{10}(\tau_{\rm jump}))$, which confirms that large relaxation times are suppressed by dynamic facilitation.}}
\label{fig:scatterplot} 
\end{figure}

We present in Fig.~\ref{fig:scatterplot} scatter plots of the energy $E_{\text{jump}}$ and the time $\tau_{\text{jump}}$ at which a trap of initial energy $E_i$ jumps. The data shown in Fig.~\ref{fig:scatterplot} are obtained for $\alpha = 1.5$ and $\beta = 3.5$ and various $\Delta$ values. The initial energies $E_i$ were drawn from the equilibrium distribution $P_\mathrm{eq}(E)$ which is centered, for these parameters, around $\overline{E} \sim 5.5$, and becomes smaller than $10^{-2}$ for $E_i$ values above 10. Starting with energies, we observe that for $\Delta = 0$, the data trivially collapse on the $E_{\text{jump}} = E_i$ line. For very small $\Delta = 10^{-4}$, the data points follow the $E_{\text{jump}} = E_i$ line at low $E_i < 6$, and depart from it above. For the majority of large $E_i$, we see that $E_{\text{jump}} < E_i$, confirming the trend of Fig.~\ref{fig:series}. With increasing $\Delta$, the data depart from the $E_{\text{jump}} = E_i$ line at smaller energies, and saturate to smaller $E_{\text{jump}}$ values at large $E_i$. 

We present in Fig.~\ref{fig:scatterplot}(b) the scatter plot of the escape time $\tau_{\text{jump}}$ from a trap of initial energy $E_i$. For $\Delta =0$, the relaxation time $\tau_{\text{jump}}$ for a given $E_i$ is Poisson-distributed with mean $e^{\beta E_i}$. As $\Delta$ increases, we see that large $\tau_{\text{jump}}$ are suppressed, with a decreasing cutoff value, mimicking the behaviour of $E_{\rm jump}$. \red{The suppression of large escape times is better appreciated in the distributions shown in the inset in Fig.~\ref{fig:scatterplot}(b).}  

The observations in Figs.~\ref{fig:series} and \ref{fig:scatterplot} qualitatively explain the shape of the relaxation spectra presented in Fig.~\ref{fig:effectdelta}. The dynamics of shallow traps is essentially unaffected by facilitation. On the contrary, the relaxation of deep traps is greatly accelerated with $\Delta >0$, as illustrated in Fig.~\ref{fig:series}(c), which effectively suppresses the long relaxation times. Facilitation thus leaves unchanged the distribution of relaxation times $G(\log\tau)$ at small $\tau$ but results in a compression and a rather sharp cutoff at large $\tau$ \red{[see the inset in Fig.~\ref{fig:scatterplot}(b)]}. At fixed frequency $\omega$, the integral in Eq.~(\ref{eq:chiG}) providing $\chi''(\omega)$ is dominated by the behaviour of the distribution $G(\log \tau)$ for $\tau\sim 1/\omega$. This explains why the spectra with and without facilitation coincide at high frequencies. Instead, at low frequencies, the compression of $G(\log\tau)$ at large times gives rise to a similar compression at low $\omega$ in the frequency domain. For example, we have $\log_{10} \tau_{\text{jump}} \sim 7$ for the largest $\Delta = 0.1$. This translates into a sharp peak around $\log_{10}\omega_p \sim -7$ in the corresponding spectrum, see Fig.~\ref{fig:effectdelta}(d).

\subsection{Approximate analytic description of the spectra}

\label{subsec:approxdynamics}

Next, we want to capture more quantitatively the effect of $\Delta$ on the distribution of timescales which is ruled by a competition between diffusion of the trap depth in energy space and thermally-activated jumps. We propose an approximate analytic treatment of the facilitated trap model, as the complete dynamics cannot be solved exactly. 

We seek an approximate description of the single-trap dynamics before it is escaped and study the diffusion of one trap in energy space before its escape. We define the probability distribution $P(E,t;E_i)$ for the energy $E$ of a single trap at time $t$, starting from the energy $E_i$ at $t=0$. This distribution follows a Fokker-Planck equation which can be derived from the master equation with the appropriate Metropolis rule:
\begin{equation}
    \begin{aligned}
    &\frac{\partial P}{\partial t}(E,t;E_i)=\\
    &D_E\frac{\partial}{\partial E}\left[-\frac{\mathrm{d}\log P_\mathrm{eq}}{\mathrm{d}E}(E)P(E,t;E_i)+\frac{\partial P}{\partial E}(E,t;E_i)\right].
    \end{aligned}
    \label{eq:pde_single}
\end{equation}
This describes the drift-diffusion of a single trap in the confining potential $V_{\rm eff}= -\log P_\mathrm{eq}$ (in units of $D_E$) and involves a diffusion constant $D_E \sim \Delta^2\Gamma_\mathrm{eq}$ (up to a numerical constant), where $\Gamma_\mathrm{eq}(T)$ represents the average jump rate at temperature $T$. The latter only depends on the equilibrium distribution and reads $\Gamma_\mathrm{eq}=\int_0^{+\infty}\mathrm{d}E P_\mathrm{eq}(E)e^{-\beta E}$. To ensure the probability conservation, we impose a hard-wall boundary condition 
\begin{equation}
    \left.\left[-\frac{\mathrm{d}\log P_\mathrm{eq}}{\mathrm{d}E}(E)P(E,t;E_i)+\frac{\partial P}{\partial E}(E,t;E_i)\right]\right\vert_{E=0}=0
    \label{eq:bc_pde}
\end{equation}
at the origin, while the initial condition is given by $P(E,t;E_i)\vert_{t=0}=\delta(E-E_i)$. 

We then come to the main approximation, which amounts to expressing the persistence function for this process as
\begin{equation}
p(t;E_i)=e^{-\int_0^t\mathrm{d}t'\Gamma(t';E_i)},
\label{eq:single_pers}
\end{equation}
with $\Gamma(t;E_i)$ the average jump rate at time $t$, given by
\begin{equation}
    \Gamma(t;E_i)=\int_0^{+\infty}P(E,t;E_i)e^{-\beta E}\mathrm{d}E.
\label{eq:gamma}
  \end{equation}
We note that the approximation in Eq.~(\ref{eq:single_pers}) yields the exact solution when $\Delta=0$, as one recovers that $p(t;E_i)=e^{-te^{-\beta E_i}}$ in this limit. 
For $\Delta>0$ it is only approximate as we have not described the full distribution of jumping times of a single trap by combining energy diffusion with Poisson-distributed relaxation times at fixed energy. Instead we have only characterised its dynamics by the average jump rate $\Gamma(t;E_i)$.

Still, our approximation captures the broad features observed in the simulations of Sec.~\ref{sec:spectra}, namely, the fact that the dynamics of shallow traps is unaffected while the one of deep traps is strongly accelerated because they diffuse towards smaller energies. Indeed, Eq.~(\ref{eq:single_pers}) yields $p(t;E_i)\sim e^{-te^{-\beta E_i}}$ in the short-time limit, and $p(t;E_i)\sim e^{-\Gamma_\mathrm{eq}t}$ in the long-time limit, because the distribution of single-trap energies tends to the equilibrium distribution $P_\mathrm{eq}$ when $t\to+\infty$, by virtue of Eq.~(\ref{eq:pde_single}). The characteristic timescale for this crossover is given by the typical diffusion timescale over an energy range of order 1, namely,
\begin{equation}
  \tau_\Delta\sim 1/D_\mathrm{E}\sim \Delta^{-2} \Gamma_{\rm eq}^{-1}(T),
\label{eq:taudelta}
\end{equation}
which depends both on $\Delta$ and $T$. 
This means that traps with $e^{\beta E_i} \ll \tau_\Delta$ are unaffected by the diffusion process and relax as when $\Delta=0$. For these traps, the typical diffusion timescale is too large and they relax by thermal activation before their energy can significantly diffuse. Instead, deep traps with $e^{\beta E_i} \gg \tau_\Delta$ have ample time to explore lower energies and thus jump faster, with a typical timescale controlled by $\tau_\Delta$. As $\tau_\Delta\sim \Delta^{-2}$, the crossover energy between these two limits becomes smaller for larger $\Delta$, in agreement with the scatter plots in Fig.~\ref{fig:scatterplot}.

\begin{figure}[t]
\includegraphics[width=\columnwidth]{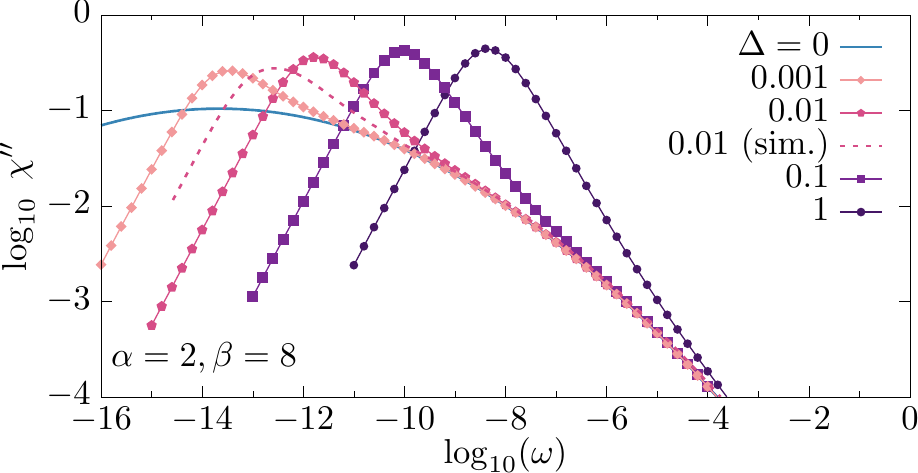}
\caption{Relaxation spectra for the facilitated Gaussian trap model ($\alpha=2$ and $\beta=8$) obtained from our approximate solution of the dynamics, solving Eqs.~(\ref{eq:persistence}, \ref{eq:chip}, \ref{eq:single_pers}). The broad feature of the direct simulations in Fig.~\ref{fig:effectdelta}(a) are correctly reproduced \red{(see the direct comparison for $\Delta=0.01$).}}
\label{fig:chi_gauss} 
\end{figure}

Moving to the average persistence in Eq.~(\ref{eq:persistence}), it follows that its time decay is unaffected by facilitation when $t \ll \tau_\Delta$ while it decays very sharply towards $0$ for $t \gg \tau_\Delta$ for $\Delta<1$. Physically, the ratio $\tau_\mathrm{eq}/\tau_\Delta$ quantifies the effective strength of the dynamic facilitation, with $\tau_\mathrm{eq}=\int_0^{+\infty}P_\mathrm{eq}(E)e^{\beta E}\mathrm{d}E$ the auto-correlation time of the $\Delta=0$ trap model. This ratio trivially increases with $\Delta$ at constant temperature. More interestingly, for a fixed value of $\Delta$, this ratio also increases with decreasing temperature. This shows that $\Delta$ itself does not uniquely characterise the strength of facilitation in the system.

To assess the quality of our approximate description of the dynamics, we solve Eq.~(\ref{eq:pde_single}) analytically for the Gaussian case $\alpha=2$ for which a closed formula for the jump rate $\Gamma(t;E_i)$ in Eq.~(\ref{eq:gamma}) can be derived, as this represents a solvable Ornstein-Uhlenbeck process in energy space~\cite{gardiner1985handbook}. Fig.~\ref{fig:chi_gauss} shows the resulting spectra for different values of $\Delta$ at inverse temperature $\beta=8$. As $\Delta$ increases, we observe the same trend as the simulation results presented in Fig.~\ref{fig:effectdelta}(a), even though the quantitative agreement is not exact, due to the approximation involved in Eq.~(\ref{eq:single_pers}).  
This suggests that our approximation correctly captures the interplay between dynamic heterogeneity and dynamic facilitation which accounts for the shape of the relaxation spectra in the facilitated trap model, and their evolution with the control parameters.

\begin{figure}[t]
\includegraphics[width=\columnwidth]{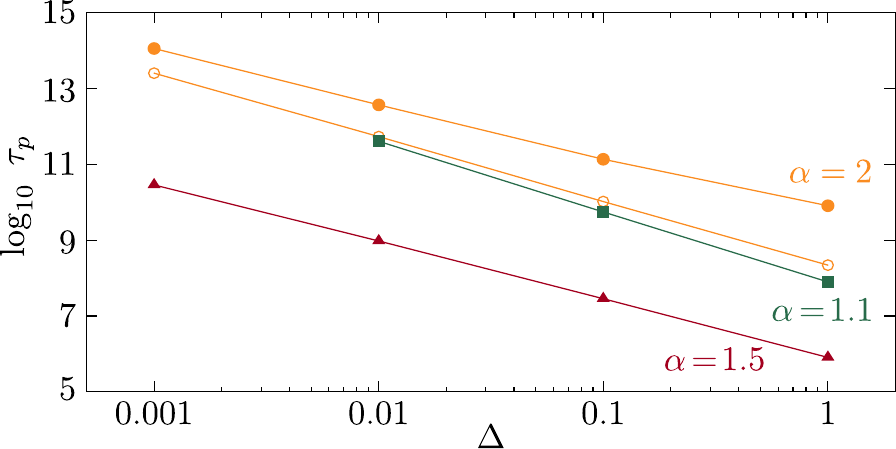}
\caption{Evolution with $\Delta$ of the relaxation time $\tau_p = 1/\omega_p$, with $\omega_p$ the location of the peak in the relaxation spectra $\chi''(\omega)$ for various sets of parameters: ($\alpha,\beta$)=(2,8), (1.5,3.5), (1.1,1.55).  Closed symbols are from direct simulations, open symbols are from the analytic approximation. The data are correctly described by an effective power law dependence: $\tau_p \sim \Delta^{-\gamma}$, with $\gamma \in \argc{1.45,1.85}$.}
\label{fig:taupdelta} 
\end{figure}

Our approximate solution reveals the existence of a diffusive timescale $\tau_\Delta$ in Eq.~(\ref{eq:taudelta}) which controls the relaxation of the deepest traps and provides a terminal cutoff to the distribution of relaxation times, which controls the linear behaviour of $\chi''(\omega)$ at low frequency. We thus analyse in Fig.~\ref{fig:taupdelta} the dependence of the relaxation time defined as $\tau_p=\omega_p^{-1}$, as a function of $\Delta$ in the simulations of the trap model. The open symbols correspond to the approximate analytic solution. In the range of $\Delta$ investigated, we find that $\tau_p \sim \Delta^{-\gamma}$ with a fitted exponent $\gamma \in \argc{1.45,1.85}$, depending on the parameters ($\alpha$, $\beta$) of the model. The measured exponent $\gamma$ increases slightly when moving from the Gaussian model $\alpha = 2$ to $\alpha =1.1$. The exponent remains slightly smaller than prediction $\gamma=2$ emerging from the diffusive timescale $\tau_\Delta$, see Eq.~(\ref{eq:taudelta}). The difference probably stems from the fact that the spectrum is an averaged quantity receiving contributions from all traps, including the ones that are less affected by facilitation. This interpretation is confirmed by our approximate analytic solution where the clear presence of the diffusive cutoff nevertheless results in an effective exponent $\gamma \sim 1.7$, different from 2.

\section{Discussion}

\label{sec:wings}

We constructed and analysed a simple version of a facilitated trap model, which combines dynamic heterogeneity and kinetic facilitation, recently identified as the two key ingredients explaining the asymmetric relaxation spectra in deeply supercooled liquids.~\cite{shortwings}. We showed that the model generates asymmetric relaxation spectra, composed of a sharp peak at low frequencies, and a much flatter and broader high-frequency signal. This shape qualitatively resembles the dielectric spectra of deeply supercooled liquids. Because there is no internal microscopic degrees of freedom inside traps, the obtained spectra lack the microscopic peak exhibited by experimental data in the THz region.

\begin{figure}[t]
\includegraphics[width=\columnwidth]{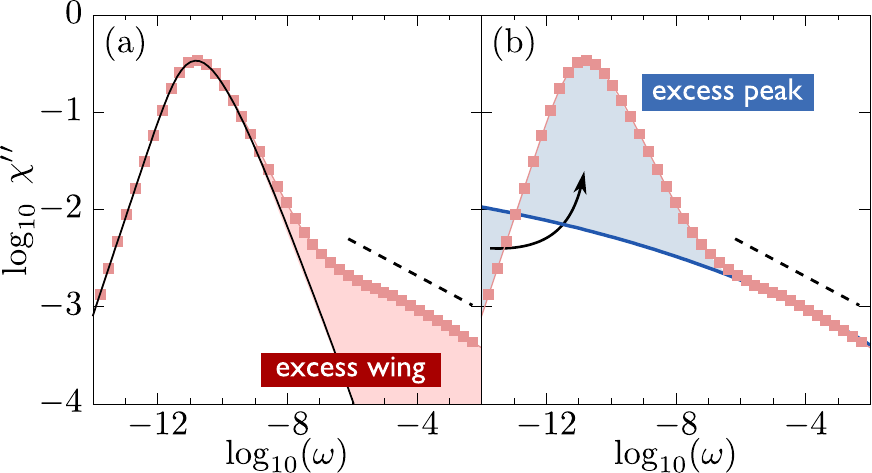}
\caption{Two interpretations of the same relaxation spectrum. (a) In the common picture, an empirical fit to the $\alpha$-peak reveals an ``excess wing'' in the high-frequency range (shaded), highlighted by the dashed line. (b) In our picture, a broad underlying distribution of timescales (blue line) becomes compressed at low frequencies by dynamic facilitation (arrow). Some of the $\alpha$-peak then appears in ``excess'' (shaded) of the intrinsic spectrum. Symbols correspond to $\alpha =1.1$, $\beta = 1.59$ and $\Delta = 0.1$.} 
\label{fig:ourpicture}
\end{figure} 

In Fig.~\ref{fig:ourpicture}(a) we reproduce the common interpretation of relaxation data generated from the present facilitated trap, where a fit to the main $\alpha$-peak reveals the existence of an additional ``excess'' signal at high frequencies -- the excess wing. Based on our results, we propose a different interpretation of the same spectrum, shown in Fig.~\ref{fig:ourpicture}(b). The self-induced heterogeneity~\cite{berthier2020self} of the supercooled liquid produces a broad distribution of activation timescales, or, equivalently, a broad spectrum (blue line). The measured spectrum does not correspond to this underlying distribution because the relaxation of the fastest regions facilitates the relaxation of the slow regions. As a result, the low-frequency part of the underlying spectrum becomes compressed in the presence of facilitation, thus producing an asymmetric shape, see Fig.~\ref{fig:ourpicture}(b). Somewhat provocatively, we might say that in this picture, the $\alpha$-peak appears in ``excess'' of a much broader underlying time distribution, while the excess wing in fact reveals the intrinsic shape of the relaxation spectrum.

\begin{figure}[t]
\includegraphics[width=\columnwidth]{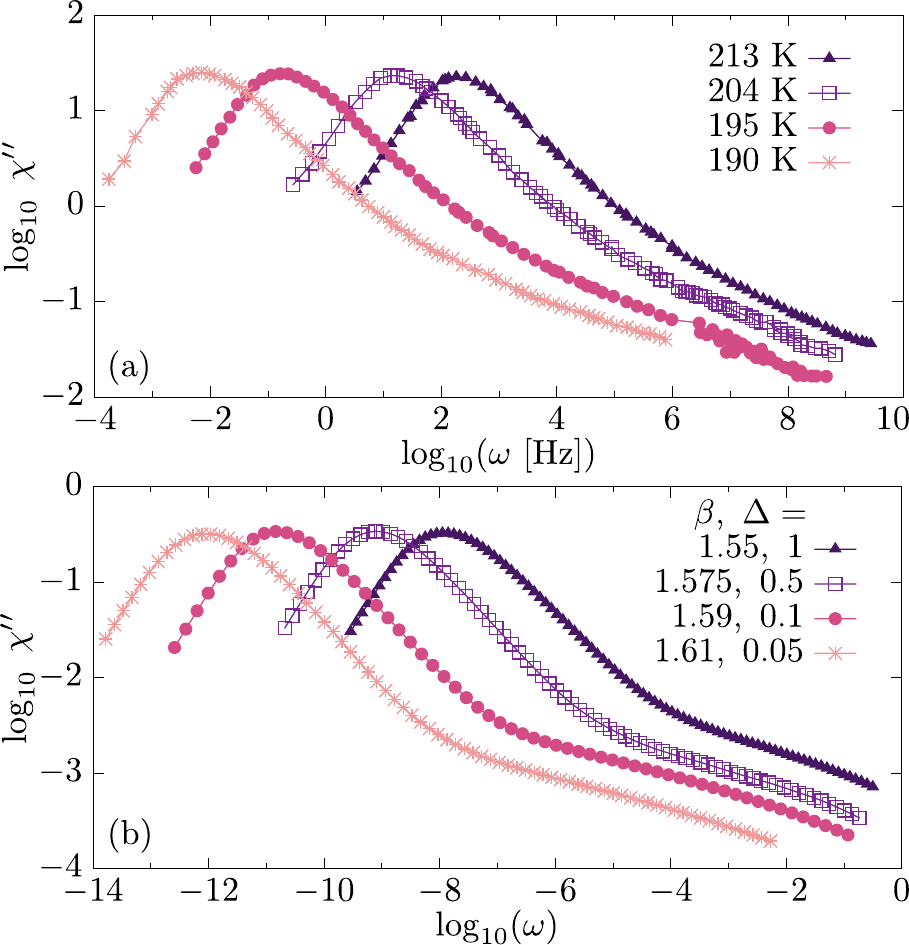}
\caption{(a) Frequency dependence of the dielectric loss (relaxation spectrum) in glycerol at various temperatures, data from Ref.~\cite{schneider2000excess}. (b) Relaxation spectra of the facilitated trap model for $\rho(E) \sim e^{-E^\alpha}$, $\alpha = 1.1$, various temperatures $1/\beta$ and values of $\Delta$. By combining dynamic heterogeneities and dynamic facilitation, the model qualitatively reproduces the experimentally-observed behaviour.} 
\label{fig:glycerol}
\end{figure} 

The facilitated trap model is an empirical model based on simple, physically-motivated, ingredients. This is not a microscopic model, it even lacks any spatial structure. As such, the scope of the model is to describe the essence of the asymmetric spectra, rather than to provide a quantitative description of any kind of data.

It is however tempting to push the model to its limits to generate relaxation spectra which resemble experimental data as closely as possible. We report in Fig.~\ref{fig:glycerol}(a) the dielectric loss of supercooled glycerol at various temperatures down to the glass transition temperature $T_g$, reproduced from Ref.~\cite{schneider2000excess}. We clearly identify a main $\alpha$-peak, which shifts to lower frequencies with decreasing temperature, as well as an ``excess wing'', which seems to flatten as the temperature decreases. We report in Fig.~\ref{fig:glycerol}(b) the relaxation spectra of the facilitated trap model for a fixed $\alpha=1.1$, obtained for various values of $\beta$ and $\Delta$. The model yields spectra which are in semi-quantitative agreement with experimental measurements. The temperature variation of the excess wing is in particular simply reproduced by changing $T$ at constant $\alpha$. The value of $\Delta$, on the other hand, was adjusted to obtain the correct location of the main $\alpha$-peak. Although the obtained $\Delta$ values decrease as $T$ decreases, the effect of facilitation, as quantified by the ratio $\tau_\mathrm{eq}/\tau_\Delta$ (see Sec.~\ref{subsec:approxdynamics}) steadily increases towards low temperatures by several orders of magnitude. Therefore, by constraining our model to yield a semi-quantitative agreement with experimental data, we conclude that the effect of dynamic facilitation becomes stronger at lower temperatures, which is fully consistent with our recent atomistic simulations~\cite{shortwings}.

There are some differences between the experimental data and results from the facilitated trap model in Fig.~\ref{fig:glycerol}. First, the excess wing obtained with the trap model is not a pure power law, as reported in experiments~\cite{nagel_scaling2} (but this is also a debated issue~\cite{schneider2000excess,lunkenheimer2002excess}). Strictly speaking, it corresponds to the high-frequency part of a very broad and very flat relaxation spectrum. However, since the observed wing extends over at most a few decades, it may be well-fitted by a power law. Another difference lies in the crossover between the relaxation peak and the high-frequency spectrum, which is quite marked in the model, while it appears much smoother in experiments. This sharp crossover may be partially explained by the lack of spatial resolution of the model, which may miss some of the heterogeneity present in real liquids. 

Our work demonstrates that a simple model combining dynamic heterogeneity and kinetic facilitation generically produces relaxation spectra that compare well with experimental and numerical findings. Since the ingredients of the model are directly motivated by a microscopic analysis of simulation results~\cite{shortwings}, we can critically revisit the assumptions made in alternative models. First, the linear superposition of two processes provides the misleading picture of distinct and independent molecular motions in the $\alpha$-peak and in the excess wing. In our model, there are not two distinct types of dynamics as the asymmetric spectrum emerges from a single type of relaxation events. Second, our approach suggests that a coupling between distinct glassy degrees of freedom is not needed to produce excess wings and complex spectra. Excess wings instead emerge in the facilitated trap model and in the atomistic simulations from a single type of glassy degree of freedom. As a consequence, excess wings appear as an intrinsic feature of slow dynamics near $T_g$. Third, our results clearly favour the interpretation of asymmetric spectra in terms of a single, asymmetric distribution of timescales. The key role played by kinetic facilitation in our argument is difficult to reconcile with the thermodynamic nature of the geometric frustration argument proposed in Ref.~\cite{viot2000heterogeneous}. Our model appears closer in spirit to the one proposed by Chamberlin~\cite{chamberlin1993non,chamberlin1998experiments,chamberlin1999mesoscopic} in which a symmetric underlying distribution (of domain sizes~\cite{chamberlin1999mesoscopic}) relaxes in a peculiar manner whereby the relaxation time of the large domains saturates to a finite limit. This is mathematically consistent with the saturation effect reported in the scatter plots in Fig.~\ref{fig:scatterplot} for the facilitated trap model, even though the kinetic argument used by Chamberlin appears physically unrelated from dynamic facilitation. Finally, our conclusions regarding the central role played by dynamic facilitation to account for excess wings are obviously consistent with results obtained with kinetically constrained models~\cite{berthier2005numerical}, and the qualitative argument proposed in Ref.~\cite{xia2001microscopic} to account for the relative narrowness of the $\alpha$-peak. \red{In all these pictures, the excess wing at high-frequency appears to result from the full decorrelation of a small fraction of the system, rather than the small motion of the entire system advocated in some NMR studies~\cite{vogel20022h}.}

In future work, it would be interesting to develop more realistic coarse-grained models of structural relaxation in supercooled liquids. An obvious step would be develop a version of the facilitated trap model with a spatial structure, to understand better the nature of the spatially heterogeneous dynamics in the presence of dynamic facilitation, in the spirit of Refs.~\cite{rehwald2010coupled,rehwald2012coupled}. It would also be interesting to characterise better the specific nature and geometry of dynamic facilitation in computer simulations~\cite{shortwings,chacko}, especially at very low temperatures that can now be simulated more easily thanks to the swap Monte Carlo algorithm.

\begin{acknowledgments}

We thank E. Bertin, G. Biroli, U. Buchenau, M. D. Ediger,  J. Kurchan, and E. R\"ossler for useful discussions and correspondance, and S. R. Nagel for providing us wth detailed explanations about the discovery of excess wings in experiments. This work was supported by a grant from the Simons Foundation (\#454933, LB), the European Research Council under the EU's Horizon 2020 Program, Grant No. 740269, a Herchel Smith Postdoctoral Research Fellowship (CS), Sidney Sussex College, Cambridge (Ramon Jenkins Research Fellowship to CS) and Capital Fund Management -- Fondation pour la Recherche (BG). 

\end{acknowledgments}

\section*{Data availability}

The data that support the findings of this study are available
from the corresponding author upon reasonable request.

\bibliography{trap.bib}

\begin{thebibliography}{91}%
\makeatletter
\providecommand \@ifxundefined [1]{%
 \@ifx{#1\undefined}
}%
\providecommand \@ifnum [1]{%
 \ifnum #1\expandafter \@firstoftwo
 \else \expandafter \@secondoftwo
 \fi
}%
\providecommand \@ifx [1]{%
 \ifx #1\expandafter \@firstoftwo
 \else \expandafter \@secondoftwo
 \fi
}%
\providecommand \natexlab [1]{#1}%
\providecommand \enquote  [1]{``#1''}%
\providecommand \bibnamefont  [1]{#1}%
\providecommand \bibfnamefont [1]{#1}%
\providecommand \citenamefont [1]{#1}%
\providecommand \href@noop [0]{\@secondoftwo}%
\providecommand \href [0]{\begingroup \@sanitize@url \@href}%
\providecommand \@href[1]{\@@startlink{#1}\@@href}%
\providecommand \@@href[1]{\endgroup#1\@@endlink}%
\providecommand \@sanitize@url [0]{\catcode `\\12\catcode `\$12\catcode
  `\&12\catcode `\#12\catcode `\^12\catcode `\_12\catcode `\%12\relax}%
\providecommand \@@startlink[1]{}%
\providecommand \@@endlink[0]{}%
\providecommand \url  [0]{\begingroup\@sanitize@url \@url }%
\providecommand \@url [1]{\endgroup\@href {#1}{\urlprefix }}%
\providecommand \urlprefix  [0]{URL }%
\providecommand \Eprint [0]{\href }%
\providecommand \doibase [0]{http://dx.doi.org/}%
\providecommand \selectlanguage [0]{\@gobble}%
\providecommand \bibinfo  [0]{\@secondoftwo}%
\providecommand \bibfield  [0]{\@secondoftwo}%
\providecommand \translation [1]{[#1]}%
\providecommand \BibitemOpen [0]{}%
\providecommand \bibitemStop [0]{}%
\providecommand \bibitemNoStop [0]{.\EOS\space}%
\providecommand \EOS [0]{\spacefactor3000\relax}%
\providecommand \BibitemShut  [1]{\csname bibitem#1\endcsname}%
\let\auto@bib@innerbib\@empty
\bibitem [{\citenamefont {Ediger}\ \emph {et~al.}(1996)\citenamefont {Ediger},
  \citenamefont {Angell},\ and\ \citenamefont {Nagel}}]{ediger1996supercooled}%
  \BibitemOpen
  \bibfield  {author} {\bibinfo {author} {\bibfnamefont {M.~D.}\ \bibnamefont
  {Ediger}}, \bibinfo {author} {\bibfnamefont {C.~A.}\ \bibnamefont {Angell}},
  \ and\ \bibinfo {author} {\bibfnamefont {S.~R.}\ \bibnamefont {Nagel}},\
  }\href {\doibase 10.1021/jp953538d} {\bibfield  {journal} {\bibinfo
  {journal} {The Journal of Physical Chemistry}\ }\textbf {\bibinfo {volume}
  {100}},\ \bibinfo {pages} {13200} (\bibinfo {year} {1996})}\BibitemShut
  {NoStop}%
\bibitem [{\citenamefont {Angell}(1995)}]{angell1995formation}%
  \BibitemOpen
  \bibfield  {author} {\bibinfo {author} {\bibfnamefont {C.~A.}\ \bibnamefont
  {Angell}},\ }\href {\doibase 10.1126/science.267.5206.1924} {\bibfield
  {journal} {\bibinfo  {journal} {Science}\ }\textbf {\bibinfo {volume}
  {267}},\ \bibinfo {pages} {1924} (\bibinfo {year} {1995})}\BibitemShut
  {NoStop}%
\bibitem [{\citenamefont {B{\"o}hmer}\ \emph {et~al.}(1993)\citenamefont
  {B{\"o}hmer}, \citenamefont {Ngai}, \citenamefont {Angell},\ and\
  \citenamefont {Plazek}}]{bohmer1993nonexponential}%
  \BibitemOpen
  \bibfield  {author} {\bibinfo {author} {\bibfnamefont {R.}~\bibnamefont
  {B{\"o}hmer}}, \bibinfo {author} {\bibfnamefont {K.~L.}\ \bibnamefont
  {Ngai}}, \bibinfo {author} {\bibfnamefont {C.~A.}\ \bibnamefont {Angell}}, \
  and\ \bibinfo {author} {\bibfnamefont {D.~J.}\ \bibnamefont {Plazek}},\
  }\href {\doibase 10.1063/1.466117} {\bibfield  {journal} {\bibinfo  {journal}
  {The Journal of Chemical Physics}\ }\textbf {\bibinfo {volume} {99}},\
  \bibinfo {pages} {4201} (\bibinfo {year} {1993})}\BibitemShut {NoStop}%
\bibitem [{\citenamefont {Lunkenheimer}\ \emph {et~al.}(2002)\citenamefont
  {Lunkenheimer}, \citenamefont {Wehn}, \citenamefont {Riegger},\ and\
  \citenamefont {Loidl}}]{lunkenheimer2002excess}%
  \BibitemOpen
  \bibfield  {author} {\bibinfo {author} {\bibfnamefont {P.}~\bibnamefont
  {Lunkenheimer}}, \bibinfo {author} {\bibfnamefont {R.}~\bibnamefont {Wehn}},
  \bibinfo {author} {\bibfnamefont {T.}~\bibnamefont {Riegger}}, \ and\
  \bibinfo {author} {\bibfnamefont {A.}~\bibnamefont {Loidl}},\ }\href
  {\doibase https://doi.org/10.1016/S0022-3093(02)01493-X} {\bibfield
  {journal} {\bibinfo  {journal} {Journal of Non-Crystalline Solids}\ }\textbf
  {\bibinfo {volume} {307-310}},\ \bibinfo {pages} {336} (\bibinfo {year}
  {2002})}\BibitemShut {NoStop}%
\bibitem [{\citenamefont {Roland}\ \emph {et~al.}(2005)\citenamefont {Roland},
  \citenamefont {Hensel-Bielowka}, \citenamefont {Paluch},\ and\ \citenamefont
  {Casalini}}]{roland2005supercooled}%
  \BibitemOpen
  \bibfield  {author} {\bibinfo {author} {\bibfnamefont {C.~M.}\ \bibnamefont
  {Roland}}, \bibinfo {author} {\bibfnamefont {S.}~\bibnamefont
  {Hensel-Bielowka}}, \bibinfo {author} {\bibfnamefont {M.}~\bibnamefont
  {Paluch}}, \ and\ \bibinfo {author} {\bibfnamefont {R.}~\bibnamefont
  {Casalini}},\ }\href {\doibase 10.1088/0034-4885/68/6/r03} {\bibfield
  {journal} {\bibinfo  {journal} {Reports on Progress in Physics}\ }\textbf
  {\bibinfo {volume} {68}},\ \bibinfo {pages} {1405} (\bibinfo {year}
  {2005})}\BibitemShut {NoStop}%
\bibitem [{\citenamefont {Ediger}(2000)}]{edigerannurev}%
  \BibitemOpen
  \bibfield  {author} {\bibinfo {author} {\bibfnamefont {M.~D.}\ \bibnamefont
  {Ediger}},\ }\href {\doibase 10.1146/annurev.physchem.51.1.99} {\bibfield
  {journal} {\bibinfo  {journal} {Annual Review of Physical Chemistry}\
  }\textbf {\bibinfo {volume} {51}},\ \bibinfo {pages} {99} (\bibinfo {year}
  {2000})}\BibitemShut {NoStop}%
\bibitem [{\citenamefont {Tarjus}(2011)}]{tarjus2011overview}%
  \BibitemOpen
  \bibfield  {author} {\bibinfo {author} {\bibfnamefont {G.}~\bibnamefont
  {Tarjus}},\ }\href@noop {} {\bibfield  {journal} {\bibinfo  {journal}
  {Dynamical Heterogeneities in Glasses, Colloids, and Granular Media}\
  }\textbf {\bibinfo {volume} {150}},\ \bibinfo {pages} {39} (\bibinfo {year}
  {2011})}\BibitemShut {NoStop}%
\bibitem [{\citenamefont {Berthier}\ and\ \citenamefont
  {Biroli}(2011)}]{berthier2011theoretical}%
  \BibitemOpen
  \bibfield  {author} {\bibinfo {author} {\bibfnamefont {L.}~\bibnamefont
  {Berthier}}\ and\ \bibinfo {author} {\bibfnamefont {G.}~\bibnamefont
  {Biroli}},\ }\href
  {https://journals.aps.org/rmp/abstract/10.1103/RevModPhys.83.587} {\bibfield
  {journal} {\bibinfo  {journal} {Reviews of modern physics}\ }\textbf
  {\bibinfo {volume} {83}},\ \bibinfo {pages} {587} (\bibinfo {year}
  {2011})}\BibitemShut {NoStop}%
\bibitem [{\citenamefont {Lunkenheimer}\ \emph {et~al.}(2000)\citenamefont
  {Lunkenheimer}, \citenamefont {Schneider}, \citenamefont {Brand},\ and\
  \citenamefont {Loid}}]{lunkenheimer2000glassy}%
  \BibitemOpen
  \bibfield  {author} {\bibinfo {author} {\bibfnamefont {P.}~\bibnamefont
  {Lunkenheimer}}, \bibinfo {author} {\bibfnamefont {U.}~\bibnamefont
  {Schneider}}, \bibinfo {author} {\bibfnamefont {R.}~\bibnamefont {Brand}}, \
  and\ \bibinfo {author} {\bibfnamefont {A.}~\bibnamefont {Loid}},\ }\href
  {https://doi.org/10.1080/001075100181259} {\bibfield  {journal} {\bibinfo
  {journal} {Contemporary Physics}\ }\textbf {\bibinfo {volume} {41}},\
  \bibinfo {pages} {15} (\bibinfo {year} {2000})}\BibitemShut {NoStop}%
\bibitem [{\citenamefont {Fl\"amig}\ \emph {et~al.}(2020)\citenamefont
  {Fl\"amig}, \citenamefont {Hofmann}, \citenamefont {Fatkullin},\ and\
  \citenamefont {R\"ossler}}]{flamig2020nmr}%
  \BibitemOpen
  \bibfield  {author} {\bibinfo {author} {\bibfnamefont {M.}~\bibnamefont
  {Fl\"amig}}, \bibinfo {author} {\bibfnamefont {M.}~\bibnamefont {Hofmann}},
  \bibinfo {author} {\bibfnamefont {N.}~\bibnamefont {Fatkullin}}, \ and\
  \bibinfo {author} {\bibfnamefont {E.}~\bibnamefont {R\"ossler}},\ }\href
  {\doibase 10.1021/acs.jpcb.9b11770} {\bibfield  {journal} {\bibinfo
  {journal} {The Journal of Physical Chemistry B}\ }\textbf {\bibinfo {volume}
  {124}},\ \bibinfo {pages} {1557} (\bibinfo {year} {2020})}\BibitemShut
  {NoStop}%
\bibitem [{\citenamefont {Schmidtke}\ \emph {et~al.}(2013)\citenamefont
  {Schmidtke}, \citenamefont {Petzold}, \citenamefont {Kahlau},\ and\
  \citenamefont {R{\"o}ssler}}]{schmidtke2013reorientational}%
  \BibitemOpen
  \bibfield  {author} {\bibinfo {author} {\bibfnamefont {B.}~\bibnamefont
  {Schmidtke}}, \bibinfo {author} {\bibfnamefont {N.}~\bibnamefont {Petzold}},
  \bibinfo {author} {\bibfnamefont {R.}~\bibnamefont {Kahlau}}, \ and\ \bibinfo
  {author} {\bibfnamefont {E.}~\bibnamefont {R{\"o}ssler}},\ }\href
  {https://doi.org/10.1063/1.4817406} {\bibfield  {journal} {\bibinfo
  {journal} {The Journal of chemical physics}\ }\textbf {\bibinfo {volume}
  {139}},\ \bibinfo {pages} {084504} (\bibinfo {year} {2013})}\BibitemShut
  {NoStop}%
\bibitem [{\citenamefont {Johari}\ and\ \citenamefont
  {Goldstein}(1970)}]{johari1970viscous}%
  \BibitemOpen
  \bibfield  {author} {\bibinfo {author} {\bibfnamefont {G.~P.}\ \bibnamefont
  {Johari}}\ and\ \bibinfo {author} {\bibfnamefont {M.}~\bibnamefont
  {Goldstein}},\ }\href {https://doi.org/10.1063/1.1674335} {\bibfield
  {journal} {\bibinfo  {journal} {The Journal of chemical physics}\ }\textbf
  {\bibinfo {volume} {53}},\ \bibinfo {pages} {2372} (\bibinfo {year}
  {1970})}\BibitemShut {NoStop}%
\bibitem [{\citenamefont {Dixon}\ \emph {et~al.}(1990)\citenamefont {Dixon},
  \citenamefont {Wu}, \citenamefont {Nagel}, \citenamefont {Williams},\ and\
  \citenamefont {Carini}}]{nagel_scaling}%
  \BibitemOpen
  \bibfield  {author} {\bibinfo {author} {\bibfnamefont {P.~K.}\ \bibnamefont
  {Dixon}}, \bibinfo {author} {\bibfnamefont {L.}~\bibnamefont {Wu}}, \bibinfo
  {author} {\bibfnamefont {S.~R.}\ \bibnamefont {Nagel}}, \bibinfo {author}
  {\bibfnamefont {B.~D.}\ \bibnamefont {Williams}}, \ and\ \bibinfo {author}
  {\bibfnamefont {J.~P.}\ \bibnamefont {Carini}},\ }\href {\doibase
  10.1103/PhysRevLett.65.1108} {\bibfield  {journal} {\bibinfo  {journal}
  {Phys. Rev. Lett.}\ }\textbf {\bibinfo {volume} {65}},\ \bibinfo {pages}
  {1108} (\bibinfo {year} {1990})}\BibitemShut {NoStop}%
\bibitem [{\citenamefont {Menon}\ and\ \citenamefont
  {Nagel}(1995)}]{nagel_scaling2}%
  \BibitemOpen
  \bibfield  {author} {\bibinfo {author} {\bibfnamefont {N.}~\bibnamefont
  {Menon}}\ and\ \bibinfo {author} {\bibfnamefont {S.~R.}\ \bibnamefont
  {Nagel}},\ }\href {\doibase 10.1103/PhysRevLett.74.1230} {\bibfield
  {journal} {\bibinfo  {journal} {Phys. Rev. Lett.}\ }\textbf {\bibinfo
  {volume} {74}},\ \bibinfo {pages} {1230} (\bibinfo {year}
  {1995})}\BibitemShut {NoStop}%
\bibitem [{\citenamefont {Menon}\ \emph {et~al.}(1992)\citenamefont {Menon},
  \citenamefont {O'Brien}, \citenamefont {Dixon}, \citenamefont {Wu},
  \citenamefont {Nagel}, \citenamefont {Williams},\ and\ \citenamefont
  {Carini}}]{menon1992wide}%
  \BibitemOpen
  \bibfield  {author} {\bibinfo {author} {\bibfnamefont {N.}~\bibnamefont
  {Menon}}, \bibinfo {author} {\bibfnamefont {K.~P.}\ \bibnamefont {O'Brien}},
  \bibinfo {author} {\bibfnamefont {P.~K.}\ \bibnamefont {Dixon}}, \bibinfo
  {author} {\bibfnamefont {L.}~\bibnamefont {Wu}}, \bibinfo {author}
  {\bibfnamefont {S.~R.}\ \bibnamefont {Nagel}}, \bibinfo {author}
  {\bibfnamefont {B.~D.}\ \bibnamefont {Williams}}, \ and\ \bibinfo {author}
  {\bibfnamefont {J.~P.}\ \bibnamefont {Carini}},\ }\href {\doibase
  https://doi.org/10.1016/S0022-3093(05)80519-8} {\bibfield  {journal}
  {\bibinfo  {journal} {Journal of Non-Crystalline Solids}\ }\textbf {\bibinfo
  {volume} {141}},\ \bibinfo {pages} {61} (\bibinfo {year} {1992})}\BibitemShut
  {NoStop}%
\bibitem [{\citenamefont {Leheny}\ and\ \citenamefont
  {Nagel}(1997)}]{leheny1997high}%
  \BibitemOpen
  \bibfield  {author} {\bibinfo {author} {\bibfnamefont {R.}~\bibnamefont
  {Leheny}}\ and\ \bibinfo {author} {\bibfnamefont {S.}~\bibnamefont {Nagel}},\
  }\href {https://doi.org/10.1209/epl/i1997-00375-2} {\bibfield  {journal}
  {\bibinfo  {journal} {EPL (Europhysics Letters)}\ }\textbf {\bibinfo {volume}
  {39}},\ \bibinfo {pages} {447} (\bibinfo {year} {1997})}\BibitemShut
  {NoStop}%
\bibitem [{\citenamefont {Leheny}\ and\ \citenamefont
  {Nagel}(1998)}]{leheny1998dielectric}%
  \BibitemOpen
  \bibfield  {author} {\bibinfo {author} {\bibfnamefont {R.~L.}\ \bibnamefont
  {Leheny}}\ and\ \bibinfo {author} {\bibfnamefont {S.~R.}\ \bibnamefont
  {Nagel}},\ }\href {\doibase https://doi.org/10.1016/S0022-3093(98)00650-4}
  {\bibfield  {journal} {\bibinfo  {journal} {Journal of Non-Crystalline
  Solids}\ }\textbf {\bibinfo {volume} {235-237}},\ \bibinfo {pages} {278}
  (\bibinfo {year} {1998})}\BibitemShut {NoStop}%
\bibitem [{\citenamefont {Adichtchev}\ \emph {et~al.}(2003)\citenamefont
  {Adichtchev}, \citenamefont {Blochowicz}, \citenamefont {Tschirwitz},
  \citenamefont {Novikov},\ and\ \citenamefont
  {R\"ossler}}]{adichtchev2003reexamination}%
  \BibitemOpen
  \bibfield  {author} {\bibinfo {author} {\bibfnamefont {S.}~\bibnamefont
  {Adichtchev}}, \bibinfo {author} {\bibfnamefont {T.}~\bibnamefont
  {Blochowicz}}, \bibinfo {author} {\bibfnamefont {C.}~\bibnamefont
  {Tschirwitz}}, \bibinfo {author} {\bibfnamefont {V.~N.}\ \bibnamefont
  {Novikov}}, \ and\ \bibinfo {author} {\bibfnamefont {E.~A.}\ \bibnamefont
  {R\"ossler}},\ }\href {\doibase 10.1103/PhysRevE.68.011504} {\bibfield
  {journal} {\bibinfo  {journal} {Phys. Rev. E}\ }\textbf {\bibinfo {volume}
  {68}},\ \bibinfo {pages} {011504} (\bibinfo {year} {2003})}\BibitemShut
  {NoStop}%
\bibitem [{\citenamefont {Blochowicz}\ \emph {et~al.}(2003)\citenamefont
  {Blochowicz}, \citenamefont {Tschirwitz}, \citenamefont {Benkhof},\ and\
  \citenamefont {R{\"o}ssler}}]{blochowicz2003susceptibility}%
  \BibitemOpen
  \bibfield  {author} {\bibinfo {author} {\bibfnamefont {T.}~\bibnamefont
  {Blochowicz}}, \bibinfo {author} {\bibfnamefont {C.}~\bibnamefont
  {Tschirwitz}}, \bibinfo {author} {\bibfnamefont {S.}~\bibnamefont {Benkhof}},
  \ and\ \bibinfo {author} {\bibfnamefont {E.}~\bibnamefont {R{\"o}ssler}},\
  }\href {https://aip.scitation.org/doi/abs/10.1063/1.1563247} {\bibfield
  {journal} {\bibinfo  {journal} {The Journal of chemical physics}\ }\textbf
  {\bibinfo {volume} {118}},\ \bibinfo {pages} {7544} (\bibinfo {year}
  {2003})}\BibitemShut {NoStop}%
\bibitem [{\citenamefont {Gainaru}\ \emph {et~al.}(2009)\citenamefont
  {Gainaru}, \citenamefont {Kahlau}, \citenamefont {R{\"o}ssler},\ and\
  \citenamefont {B{\"o}hmer}}]{gainaru2009evolution}%
  \BibitemOpen
  \bibfield  {author} {\bibinfo {author} {\bibfnamefont {C.}~\bibnamefont
  {Gainaru}}, \bibinfo {author} {\bibfnamefont {R.}~\bibnamefont {Kahlau}},
  \bibinfo {author} {\bibfnamefont {E.~A.}\ \bibnamefont {R{\"o}ssler}}, \ and\
  \bibinfo {author} {\bibfnamefont {R.}~\bibnamefont {B{\"o}hmer}},\ }\href
  {https://doi.org/10.1063/1.3258430} {\bibfield  {journal} {\bibinfo
  {journal} {The Journal of chemical physics}\ }\textbf {\bibinfo {volume}
  {131}},\ \bibinfo {pages} {184510} (\bibinfo {year} {2009})}\BibitemShut
  {NoStop}%
\bibitem [{\citenamefont {Caporaletti}\ \emph {et~al.}(2021)\citenamefont
  {Caporaletti}, \citenamefont {Capaccioli}, \citenamefont {Valenti},
  \citenamefont {Mikolasek}, \citenamefont {Chumakov},\ and\ \citenamefont
  {Monaco}}]{caporaletti2021experimental}%
  \BibitemOpen
  \bibfield  {author} {\bibinfo {author} {\bibfnamefont {F.}~\bibnamefont
  {Caporaletti}}, \bibinfo {author} {\bibfnamefont {S.}~\bibnamefont
  {Capaccioli}}, \bibinfo {author} {\bibfnamefont {S.}~\bibnamefont {Valenti}},
  \bibinfo {author} {\bibfnamefont {M.}~\bibnamefont {Mikolasek}}, \bibinfo
  {author} {\bibfnamefont {A.}~\bibnamefont {Chumakov}}, \ and\ \bibinfo
  {author} {\bibfnamefont {G.}~\bibnamefont {Monaco}},\ }\href
  {https://www.nature.com/articles/s41467-021-22154-8} {\bibfield  {journal}
  {\bibinfo  {journal} {Nature communications}\ }\textbf {\bibinfo {volume}
  {12}},\ \bibinfo {pages} {1} (\bibinfo {year} {2021})}\BibitemShut {NoStop}%
\bibitem [{\citenamefont {Blochowicz}\ \emph {et~al.}(2006)\citenamefont
  {Blochowicz}, \citenamefont {Brodin},\ and\ \citenamefont
  {R{\"o}ssler}}]{blochowicz2006evolution}%
  \BibitemOpen
  \bibfield  {author} {\bibinfo {author} {\bibfnamefont {T.}~\bibnamefont
  {Blochowicz}}, \bibinfo {author} {\bibfnamefont {A.}~\bibnamefont {Brodin}},
  \ and\ \bibinfo {author} {\bibfnamefont {E.~A.}\ \bibnamefont
  {R{\"o}ssler}},\ }\href@noop {} {\bibfield  {journal} {\bibinfo  {journal}
  {Fractals, Diffusion, and Relaxation in Disordered Complex Systems: Advances
  in Chemical Physics, Part A}\ }\textbf {\bibinfo {volume} {133}},\ \bibinfo
  {pages} {127} (\bibinfo {year} {2006})}\BibitemShut {NoStop}%
\bibitem [{\citenamefont {Gainaru}(2019)}]{gainaru2019spectral}%
  \BibitemOpen
  \bibfield  {author} {\bibinfo {author} {\bibfnamefont {C.}~\bibnamefont
  {Gainaru}},\ }\href@noop {} {\bibfield  {journal} {\bibinfo  {journal}
  {Physical Review E}\ }\textbf {\bibinfo {volume} {100}},\ \bibinfo {pages}
  {020601} (\bibinfo {year} {2019})}\BibitemShut {NoStop}%
\bibitem [{\citenamefont {Ngai}\ and\ \citenamefont
  {Paluch}(2004)}]{ngai_classification}%
  \BibitemOpen
  \bibfield  {author} {\bibinfo {author} {\bibfnamefont {K.~L.}\ \bibnamefont
  {Ngai}}\ and\ \bibinfo {author} {\bibfnamefont {M.}~\bibnamefont {Paluch}},\
  }\href {\doibase 10.1063/1.1630295} {\bibfield  {journal} {\bibinfo
  {journal} {The Journal of Chemical Physics}\ }\textbf {\bibinfo {volume}
  {120}},\ \bibinfo {pages} {857} (\bibinfo {year} {2004})}\BibitemShut
  {NoStop}%
\bibitem [{\citenamefont {Wu}(1991)}]{wu1991relaxation}%
  \BibitemOpen
  \bibfield  {author} {\bibinfo {author} {\bibfnamefont {L.}~\bibnamefont
  {Wu}},\ }\href {\doibase 10.1103/PhysRevB.43.9906} {\bibfield  {journal}
  {\bibinfo  {journal} {Phys. Rev. B}\ }\textbf {\bibinfo {volume} {43}},\
  \bibinfo {pages} {9906} (\bibinfo {year} {1991})}\BibitemShut {NoStop}%
\bibitem [{\citenamefont {Vogel}\ \emph {et~al.}(2002)\citenamefont {Vogel},
  \citenamefont {Tschirwitz}, \citenamefont {Schneider}, \citenamefont
  {Koplin}, \citenamefont {Medick},\ and\ \citenamefont
  {R{\"o}ssler}}]{vogel20022h}%
  \BibitemOpen
  \bibfield  {author} {\bibinfo {author} {\bibfnamefont {M.}~\bibnamefont
  {Vogel}}, \bibinfo {author} {\bibfnamefont {C.}~\bibnamefont {Tschirwitz}},
  \bibinfo {author} {\bibfnamefont {G.}~\bibnamefont {Schneider}}, \bibinfo
  {author} {\bibfnamefont {C.}~\bibnamefont {Koplin}}, \bibinfo {author}
  {\bibfnamefont {P.}~\bibnamefont {Medick}}, \ and\ \bibinfo {author}
  {\bibfnamefont {E.}~\bibnamefont {R{\"o}ssler}},\ }\href@noop {} {\bibfield
  {journal} {\bibinfo  {journal} {Journal of non-crystalline solids}\ }\textbf
  {\bibinfo {volume} {307}},\ \bibinfo {pages} {326} (\bibinfo {year}
  {2002})}\BibitemShut {NoStop}%
\bibitem [{\citenamefont {Petzold}\ \emph {et~al.}(2013)\citenamefont
  {Petzold}, \citenamefont {Schmidtke}, \citenamefont {Kahlau}, \citenamefont
  {Bock}, \citenamefont {Meier}, \citenamefont {Micko}, \citenamefont {Kruk},\
  and\ \citenamefont {R{\"o}ssler}}]{petzold2013evolution}%
  \BibitemOpen
  \bibfield  {author} {\bibinfo {author} {\bibfnamefont {N.}~\bibnamefont
  {Petzold}}, \bibinfo {author} {\bibfnamefont {B.}~\bibnamefont {Schmidtke}},
  \bibinfo {author} {\bibfnamefont {R.}~\bibnamefont {Kahlau}}, \bibinfo
  {author} {\bibfnamefont {D.}~\bibnamefont {Bock}}, \bibinfo {author}
  {\bibfnamefont {R.}~\bibnamefont {Meier}}, \bibinfo {author} {\bibfnamefont
  {B.}~\bibnamefont {Micko}}, \bibinfo {author} {\bibfnamefont
  {D.}~\bibnamefont {Kruk}}, \ and\ \bibinfo {author} {\bibfnamefont
  {E.}~\bibnamefont {R{\"o}ssler}},\ }\href {https://doi.org/10.1063/1.4770055}
  {\bibfield  {journal} {\bibinfo  {journal} {The Journal of chemical physics}\
  }\textbf {\bibinfo {volume} {138}},\ \bibinfo {pages} {12A510} (\bibinfo
  {year} {2013})}\BibitemShut {NoStop}%
\bibitem [{\citenamefont {K{\"o}rber}\ \emph {et~al.}(2020)\citenamefont
  {K{\"o}rber}, \citenamefont {St{\"a}glich}, \citenamefont {Gainaru},
  \citenamefont {B{\"o}hmer},\ and\ \citenamefont
  {R{\"o}ssler}}]{korber2020systematic}%
  \BibitemOpen
  \bibfield  {author} {\bibinfo {author} {\bibfnamefont {T.}~\bibnamefont
  {K{\"o}rber}}, \bibinfo {author} {\bibfnamefont {R.}~\bibnamefont
  {St{\"a}glich}}, \bibinfo {author} {\bibfnamefont {C.}~\bibnamefont
  {Gainaru}}, \bibinfo {author} {\bibfnamefont {R.}~\bibnamefont {B{\"o}hmer}},
  \ and\ \bibinfo {author} {\bibfnamefont {E.~A.}\ \bibnamefont
  {R{\"o}ssler}},\ }\href {https://doi.org/10.1063/5.0022155} {\bibfield
  {journal} {\bibinfo  {journal} {The Journal of Chemical Physics}\ }\textbf
  {\bibinfo {volume} {153}},\ \bibinfo {pages} {124510} (\bibinfo {year}
  {2020})}\BibitemShut {NoStop}%
\bibitem [{\citenamefont {Buchenau}\ \emph {et~al.}(2020)\citenamefont
  {Buchenau}, \citenamefont {D'Angelo}, \citenamefont {Carini}, \citenamefont
  {Liu},\ and\ \citenamefont {Ramos}}]{buchenau2020sound}%
  \BibitemOpen
  \bibfield  {author} {\bibinfo {author} {\bibfnamefont {U.}~\bibnamefont
  {Buchenau}}, \bibinfo {author} {\bibfnamefont {G.}~\bibnamefont {D'Angelo}},
  \bibinfo {author} {\bibfnamefont {G.}~\bibnamefont {Carini}}, \bibinfo
  {author} {\bibfnamefont {X.}~\bibnamefont {Liu}}, \ and\ \bibinfo {author}
  {\bibfnamefont {M.}~\bibnamefont {Ramos}},\ }\href
  {https://arxiv.org/abs/2012.10139} {\bibfield  {journal} {\bibinfo  {journal}
  {arXiv preprint arXiv:2012.10139}\ } (\bibinfo {year} {2020})}\BibitemShut
  {NoStop}%
\bibitem [{\citenamefont {Kudlik}\ \emph {et~al.}(1998)\citenamefont {Kudlik},
  \citenamefont {Tschirwitz}, \citenamefont {Blochowicz}, \citenamefont
  {Benkhof},\ and\ \citenamefont {R{\"o}ssler}}]{kudlik1998slow}%
  \BibitemOpen
  \bibfield  {author} {\bibinfo {author} {\bibfnamefont {A.}~\bibnamefont
  {Kudlik}}, \bibinfo {author} {\bibfnamefont {C.}~\bibnamefont {Tschirwitz}},
  \bibinfo {author} {\bibfnamefont {T.}~\bibnamefont {Blochowicz}}, \bibinfo
  {author} {\bibfnamefont {S.}~\bibnamefont {Benkhof}}, \ and\ \bibinfo
  {author} {\bibfnamefont {E.}~\bibnamefont {R{\"o}ssler}},\ }\href
  {https://www.sciencedirect.com/science/article/pii/S0022309398005109}
  {\bibfield  {journal} {\bibinfo  {journal} {Journal of non-crystalline
  solids}\ }\textbf {\bibinfo {volume} {235}},\ \bibinfo {pages} {406}
  (\bibinfo {year} {1998})}\BibitemShut {NoStop}%
\bibitem [{\citenamefont {Gainaru}\ \emph {et~al.}(2010)\citenamefont
  {Gainaru}, \citenamefont {B{\"o}hmer}, \citenamefont {Kahlau},\ and\
  \citenamefont {R{\"o}ssler}}]{gainaru2010energy}%
  \BibitemOpen
  \bibfield  {author} {\bibinfo {author} {\bibfnamefont {C.}~\bibnamefont
  {Gainaru}}, \bibinfo {author} {\bibfnamefont {R.}~\bibnamefont {B{\"o}hmer}},
  \bibinfo {author} {\bibfnamefont {R.}~\bibnamefont {Kahlau}}, \ and\ \bibinfo
  {author} {\bibfnamefont {E.}~\bibnamefont {R{\"o}ssler}},\ }\href {\doibase
  10.1103/PhysRevB.82.104205} {\bibfield  {journal} {\bibinfo  {journal}
  {Physical Review B}\ }\textbf {\bibinfo {volume} {82}},\ \bibinfo {pages}
  {104205} (\bibinfo {year} {2010})}\BibitemShut {NoStop}%
\bibitem [{\citenamefont {Diezemann}\ \emph {et~al.}(1999)\citenamefont
  {Diezemann}, \citenamefont {Mohanty},\ and\ \citenamefont
  {Oppenheim}}]{diezemann1999slow}%
  \BibitemOpen
  \bibfield  {author} {\bibinfo {author} {\bibfnamefont {G.}~\bibnamefont
  {Diezemann}}, \bibinfo {author} {\bibfnamefont {U.}~\bibnamefont {Mohanty}},
  \ and\ \bibinfo {author} {\bibfnamefont {I.}~\bibnamefont {Oppenheim}},\
  }\href {\doibase 10.1103/PhysRevE.59.2067} {\bibfield  {journal} {\bibinfo
  {journal} {Phys. Rev. E}\ }\textbf {\bibinfo {volume} {59}},\ \bibinfo
  {pages} {2067} (\bibinfo {year} {1999})}\BibitemShut {NoStop}%
\bibitem [{\citenamefont {Mohanty}\ \emph {et~al.}(2000)\citenamefont
  {Mohanty}, \citenamefont {Diezemann},\ and\ \citenamefont
  {Oppenheim}}]{mohanty2000nature}%
  \BibitemOpen
  \bibfield  {author} {\bibinfo {author} {\bibfnamefont {U.}~\bibnamefont
  {Mohanty}}, \bibinfo {author} {\bibfnamefont {G.}~\bibnamefont {Diezemann}},
  \ and\ \bibinfo {author} {\bibfnamefont {I.}~\bibnamefont {Oppenheim}},\
  }\href {\doibase 10.1088/0953-8984/12/29/315} {\bibfield  {journal} {\bibinfo
   {journal} {Journal of Physics: Condensed Matter}\ }\textbf {\bibinfo
  {volume} {12}},\ \bibinfo {pages} {6431} (\bibinfo {year}
  {2000})}\BibitemShut {NoStop}%
\bibitem [{\citenamefont {G\"otze}\ and\ \citenamefont
  {Sj\"ogren}(1989)}]{gotze1989beta}%
  \BibitemOpen
  \bibfield  {author} {\bibinfo {author} {\bibfnamefont {W.}~\bibnamefont
  {G\"otze}}\ and\ \bibinfo {author} {\bibfnamefont {L.}~\bibnamefont
  {Sj\"ogren}},\ }\href {\doibase 10.1088/0953-8984/1/26/014} {\bibfield
  {journal} {\bibinfo  {journal} {Journal of Physics: Condensed Matter}\
  }\textbf {\bibinfo {volume} {1}},\ \bibinfo {pages} {4183} (\bibinfo {year}
  {1989})}\BibitemShut {NoStop}%
\bibitem [{\citenamefont {G\"otze}\ and\ \citenamefont
  {Sperl}(2002)}]{gotze2002logarithmic}%
  \BibitemOpen
  \bibfield  {author} {\bibinfo {author} {\bibfnamefont {W.}~\bibnamefont
  {G\"otze}}\ and\ \bibinfo {author} {\bibfnamefont {M.}~\bibnamefont
  {Sperl}},\ }\href {\doibase 10.1103/PhysRevE.66.011405} {\bibfield  {journal}
  {\bibinfo  {journal} {Phys. Rev. E}\ }\textbf {\bibinfo {volume} {66}},\
  \bibinfo {pages} {011405} (\bibinfo {year} {2002})}\BibitemShut {NoStop}%
\bibitem [{\citenamefont {Domschke}\ \emph {et~al.}(2011)\citenamefont
  {Domschke}, \citenamefont {Marsilius}, \citenamefont {Blochowicz},\ and\
  \citenamefont {Voigtmann}}]{domschke2011glassy}%
  \BibitemOpen
  \bibfield  {author} {\bibinfo {author} {\bibfnamefont {M.}~\bibnamefont
  {Domschke}}, \bibinfo {author} {\bibfnamefont {M.}~\bibnamefont {Marsilius}},
  \bibinfo {author} {\bibfnamefont {T.}~\bibnamefont {Blochowicz}}, \ and\
  \bibinfo {author} {\bibfnamefont {T.}~\bibnamefont {Voigtmann}},\ }\href
  {\doibase 10.1103/PhysRevE.84.031506} {\bibfield  {journal} {\bibinfo
  {journal} {Phys. Rev. E}\ }\textbf {\bibinfo {volume} {84}},\ \bibinfo
  {pages} {031506} (\bibinfo {year} {2011})}\BibitemShut {NoStop}%
\bibitem [{\citenamefont {Chong}\ and\ \citenamefont
  {G\"otze}(2002)}]{chong2002structural}%
  \BibitemOpen
  \bibfield  {author} {\bibinfo {author} {\bibfnamefont {S.-H.}\ \bibnamefont
  {Chong}}\ and\ \bibinfo {author} {\bibfnamefont {W.}~\bibnamefont
  {G\"otze}},\ }\href {\doibase 10.1103/PhysRevE.65.051201} {\bibfield
  {journal} {\bibinfo  {journal} {Phys. Rev. E}\ }\textbf {\bibinfo {volume}
  {65}},\ \bibinfo {pages} {051201} (\bibinfo {year} {2002})}\BibitemShut
  {NoStop}%
\bibitem [{\citenamefont {Cummins}(2005)}]{cummins2005dynamics}%
  \BibitemOpen
  \bibfield  {author} {\bibinfo {author} {\bibfnamefont {H.}~\bibnamefont
  {Cummins}},\ }\href {\doibase 10.1088/0953-8984/17/10/003} {\bibfield
  {journal} {\bibinfo  {journal} {Journal of Physics: Condensed Matter}\
  }\textbf {\bibinfo {volume} {17}},\ \bibinfo {pages} {1457} (\bibinfo {year}
  {2005})}\BibitemShut {NoStop}%
\bibitem [{\citenamefont {Sj{\"o}gren}(1986)}]{sjogren1986diffusion}%
  \BibitemOpen
  \bibfield  {author} {\bibinfo {author} {\bibfnamefont {L.}~\bibnamefont
  {Sj{\"o}gren}},\ }\href {\doibase 10.1103/PhysRevA.33.1254} {\bibfield
  {journal} {\bibinfo  {journal} {Physical Review A}\ }\textbf {\bibinfo
  {volume} {33}},\ \bibinfo {pages} {1254} (\bibinfo {year}
  {1986})}\BibitemShut {NoStop}%
\bibitem [{\citenamefont {Crisanti}\ \emph {et~al.}(2011)\citenamefont
  {Crisanti}, \citenamefont {Leuzzi},\ and\ \citenamefont
  {Paoluzzi}}]{crisanti2011statistical}%
  \BibitemOpen
  \bibfield  {author} {\bibinfo {author} {\bibfnamefont {A.}~\bibnamefont
  {Crisanti}}, \bibinfo {author} {\bibfnamefont {L.}~\bibnamefont {Leuzzi}}, \
  and\ \bibinfo {author} {\bibfnamefont {M.}~\bibnamefont {Paoluzzi}},\ }\href
  {https://link.springer.com/article/10.1140/epje/i2011-11098-3} {\bibfield
  {journal} {\bibinfo  {journal} {The European Physical Journal E}\ }\textbf
  {\bibinfo {volume} {34}},\ \bibinfo {pages} {1} (\bibinfo {year}
  {2011})}\BibitemShut {NoStop}%
\bibitem [{\citenamefont {Crisanti}\ and\ \citenamefont
  {Leuzzi}(2015)}]{crisanti2015simple}%
  \BibitemOpen
  \bibfield  {author} {\bibinfo {author} {\bibfnamefont {A.}~\bibnamefont
  {Crisanti}}\ and\ \bibinfo {author} {\bibfnamefont {L.}~\bibnamefont
  {Leuzzi}},\ }\href {\doibase
  https://doi.org/10.1016/j.jnoncrysol.2014.07.048} {\bibfield  {journal}
  {\bibinfo  {journal} {Journal of Non-Crystalline Solids}\ }\textbf {\bibinfo
  {volume} {407}},\ \bibinfo {pages} {110} (\bibinfo {year}
  {2015})}\BibitemShut {NoStop}%
\bibitem [{\citenamefont {Sethna}\ \emph {et~al.}(1991)\citenamefont {Sethna},
  \citenamefont {Shore},\ and\ \citenamefont {Huang}}]{sethna1991scaling}%
  \BibitemOpen
  \bibfield  {author} {\bibinfo {author} {\bibfnamefont {J.~P.}\ \bibnamefont
  {Sethna}}, \bibinfo {author} {\bibfnamefont {J.~D.}\ \bibnamefont {Shore}}, \
  and\ \bibinfo {author} {\bibfnamefont {M.}~\bibnamefont {Huang}},\ }\href
  {\doibase 10.1103/PhysRevB.44.4943} {\bibfield  {journal} {\bibinfo
  {journal} {Phys. Rev. B}\ }\textbf {\bibinfo {volume} {44}},\ \bibinfo
  {pages} {4943} (\bibinfo {year} {1991})}\BibitemShut {NoStop}%
\bibitem [{\citenamefont {Stevenson}\ and\ \citenamefont
  {Wolynes}(2010)}]{stevenson2010universal}%
  \BibitemOpen
  \bibfield  {author} {\bibinfo {author} {\bibfnamefont {J.~D.}\ \bibnamefont
  {Stevenson}}\ and\ \bibinfo {author} {\bibfnamefont {P.~G.}\ \bibnamefont
  {Wolynes}},\ }\href {https://www.nature.com/articles/nphys1432} {\bibfield
  {journal} {\bibinfo  {journal} {Nature physics}\ }\textbf {\bibinfo {volume}
  {6}},\ \bibinfo {pages} {62} (\bibinfo {year} {2010})}\BibitemShut {NoStop}%
\bibitem [{\citenamefont {Viot}\ \emph {et~al.}(2000)\citenamefont {Viot},
  \citenamefont {Tarjus},\ and\ \citenamefont
  {Kivelson}}]{viot2000heterogeneous}%
  \BibitemOpen
  \bibfield  {author} {\bibinfo {author} {\bibfnamefont {P.}~\bibnamefont
  {Viot}}, \bibinfo {author} {\bibfnamefont {G.}~\bibnamefont {Tarjus}}, \ and\
  \bibinfo {author} {\bibfnamefont {D.}~\bibnamefont {Kivelson}},\ }\href
  {https://doi.org/10.1063/1.481674} {\bibfield  {journal} {\bibinfo  {journal}
  {The Journal of Chemical Physics}\ }\textbf {\bibinfo {volume} {112}},\
  \bibinfo {pages} {10368} (\bibinfo {year} {2000})}\BibitemShut {NoStop}%
\bibitem [{\citenamefont {Chamberlin}(1999)}]{chamberlin1999mesoscopic}%
  \BibitemOpen
  \bibfield  {author} {\bibinfo {author} {\bibfnamefont {R.~V.}\ \bibnamefont
  {Chamberlin}},\ }\href {\doibase 10.1103/PhysRevLett.82.2520} {\bibfield
  {journal} {\bibinfo  {journal} {Phys. Rev. Lett.}\ }\textbf {\bibinfo
  {volume} {82}},\ \bibinfo {pages} {2520} (\bibinfo {year}
  {1999})}\BibitemShut {NoStop}%
\bibitem [{\citenamefont {Dyre}\ and\ \citenamefont
  {Schr\o{}der}(2000)}]{dyre2000universality}%
  \BibitemOpen
  \bibfield  {author} {\bibinfo {author} {\bibfnamefont {J.~C.}\ \bibnamefont
  {Dyre}}\ and\ \bibinfo {author} {\bibfnamefont {T.~B.}\ \bibnamefont
  {Schr\o{}der}},\ }\href {\doibase 10.1103/RevModPhys.72.873} {\bibfield
  {journal} {\bibinfo  {journal} {Rev. Mod. Phys.}\ }\textbf {\bibinfo {volume}
  {72}},\ \bibinfo {pages} {873} (\bibinfo {year} {2000})}\BibitemShut
  {NoStop}%
\bibitem [{\citenamefont {Berthier}\ and\ \citenamefont
  {Garrahan}(2005)}]{berthier2005numerical}%
  \BibitemOpen
  \bibfield  {author} {\bibinfo {author} {\bibfnamefont {L.}~\bibnamefont
  {Berthier}}\ and\ \bibinfo {author} {\bibfnamefont {J.~P.}\ \bibnamefont
  {Garrahan}},\ }\href {\doibase 10.1021/jp045491e} {\bibfield  {journal}
  {\bibinfo  {journal} {The Journal of Physical Chemistry B}\ }\textbf
  {\bibinfo {volume} {109}},\ \bibinfo {pages} {3578} (\bibinfo {year}
  {2005})}\BibitemShut {NoStop}%
\bibitem [{\citenamefont {Tarjus}\ \emph {et~al.}(2005)\citenamefont {Tarjus},
  \citenamefont {Kivelson}, \citenamefont {Nussinov},\ and\ \citenamefont
  {Viot}}]{tarjus2005frustration}%
  \BibitemOpen
  \bibfield  {author} {\bibinfo {author} {\bibfnamefont {G.}~\bibnamefont
  {Tarjus}}, \bibinfo {author} {\bibfnamefont {S.~A.}\ \bibnamefont
  {Kivelson}}, \bibinfo {author} {\bibfnamefont {Z.}~\bibnamefont {Nussinov}},
  \ and\ \bibinfo {author} {\bibfnamefont {P.}~\bibnamefont {Viot}},\ }\href
  {\doibase 10.1088/0953-8984/17/50/r01} {\bibfield  {journal} {\bibinfo
  {journal} {Journal of Physics: Condensed Matter}\ }\textbf {\bibinfo {volume}
  {17}},\ \bibinfo {pages} {R1143} (\bibinfo {year} {2005})}\BibitemShut
  {NoStop}%
\bibitem [{\citenamefont {Chamberlin}(1993)}]{chamberlin1993non}%
  \BibitemOpen
  \bibfield  {author} {\bibinfo {author} {\bibfnamefont {R.~V.}\ \bibnamefont
  {Chamberlin}},\ }\href {\doibase 10.1103/PhysRevB.48.15638} {\bibfield
  {journal} {\bibinfo  {journal} {Phys. Rev. B}\ }\textbf {\bibinfo {volume}
  {48}},\ \bibinfo {pages} {15638} (\bibinfo {year} {1993})}\BibitemShut
  {NoStop}%
\bibitem [{\citenamefont {Chamberlin}(1998)}]{chamberlin1998experiments}%
  \BibitemOpen
  \bibfield  {author} {\bibinfo {author} {\bibfnamefont {R.~V.}\ \bibnamefont
  {Chamberlin}},\ }\href {\doibase 10.1080/01411599808209287} {\bibfield
  {journal} {\bibinfo  {journal} {Phase Transitions: A Multinational Journal}\
  }\textbf {\bibinfo {volume} {65}},\ \bibinfo {pages} {169} (\bibinfo {year}
  {1998})}\BibitemShut {NoStop}%
\bibitem [{\citenamefont {Garrahan}\ \emph {et~al.}(2011)\citenamefont
  {Garrahan}, \citenamefont {Sollich},\ and\ \citenamefont
  {Toninelli}}]{garrahan2011kinetically}%
  \BibitemOpen
  \bibfield  {author} {\bibinfo {author} {\bibfnamefont {J.~P.}\ \bibnamefont
  {Garrahan}}, \bibinfo {author} {\bibfnamefont {P.}~\bibnamefont {Sollich}}, \
  and\ \bibinfo {author} {\bibfnamefont {C.}~\bibnamefont {Toninelli}},\
  }\href@noop {} {\bibfield  {journal} {\bibinfo  {journal} {Dynamical
  heterogeneities in glasses, colloids, and granular media}\ }\textbf {\bibinfo
  {volume} {150}},\ \bibinfo {pages} {111} (\bibinfo {year}
  {2011})}\BibitemShut {NoStop}%
\bibitem [{\citenamefont {Chandler}\ and\ \citenamefont
  {Garrahan}(2010)}]{chandler2010dynamics}%
  \BibitemOpen
  \bibfield  {author} {\bibinfo {author} {\bibfnamefont {D.}~\bibnamefont
  {Chandler}}\ and\ \bibinfo {author} {\bibfnamefont {J.~P.}\ \bibnamefont
  {Garrahan}},\ }\href {\doibase 10.1146/annurev.physchem.040808.090405}
  {\bibfield  {journal} {\bibinfo  {journal} {Annual Review of Physical
  Chemistry}\ }\textbf {\bibinfo {volume} {61}},\ \bibinfo {pages} {191}
  (\bibinfo {year} {2010})}\BibitemShut {NoStop}%
\bibitem [{\citenamefont {Kirkpatrick}\ \emph {et~al.}(1989)\citenamefont
  {Kirkpatrick}, \citenamefont {Thirumalai},\ and\ \citenamefont
  {Wolynes}}]{kirkpatrick1989scaling}%
  \BibitemOpen
  \bibfield  {author} {\bibinfo {author} {\bibfnamefont {T.~R.}\ \bibnamefont
  {Kirkpatrick}}, \bibinfo {author} {\bibfnamefont {D.}~\bibnamefont
  {Thirumalai}}, \ and\ \bibinfo {author} {\bibfnamefont {P.~G.}\ \bibnamefont
  {Wolynes}},\ }\href {\doibase 10.1103/PhysRevA.40.1045} {\bibfield  {journal}
  {\bibinfo  {journal} {Phys. Rev. A}\ }\textbf {\bibinfo {volume} {40}},\
  \bibinfo {pages} {1045} (\bibinfo {year} {1989})}\BibitemShut {NoStop}%
\bibitem [{\citenamefont {Lubchenko}\ and\ \citenamefont
  {Wolynes}(2007)}]{lubchenko2007theory}%
  \BibitemOpen
  \bibfield  {author} {\bibinfo {author} {\bibfnamefont {V.}~\bibnamefont
  {Lubchenko}}\ and\ \bibinfo {author} {\bibfnamefont {P.~G.}\ \bibnamefont
  {Wolynes}},\ }\href {\doibase 10.1146/annurev.physchem.58.032806.104653}
  {\bibfield  {journal} {\bibinfo  {journal} {Annual Review of Physical
  Chemistry}\ }\textbf {\bibinfo {volume} {58}},\ \bibinfo {pages} {235}
  (\bibinfo {year} {2007})}\BibitemShut {NoStop}%
\bibitem [{\citenamefont {Xia}\ and\ \citenamefont
  {Wolynes}(2001)}]{xia2001microscopic}%
  \BibitemOpen
  \bibfield  {author} {\bibinfo {author} {\bibfnamefont {X.}~\bibnamefont
  {Xia}}\ and\ \bibinfo {author} {\bibfnamefont {P.~G.}\ \bibnamefont
  {Wolynes}},\ }\href {\doibase 10.1103/PhysRevLett.86.5526} {\bibfield
  {journal} {\bibinfo  {journal} {Phys. Rev. Lett.}\ }\textbf {\bibinfo
  {volume} {86}},\ \bibinfo {pages} {5526} (\bibinfo {year}
  {2001})}\BibitemShut {NoStop}%
\bibitem [{\citenamefont {Bhattacharyya}\ \emph {et~al.}(2008)\citenamefont
  {Bhattacharyya}, \citenamefont {Bagchi},\ and\ \citenamefont
  {Wolynes}}]{bhattacharyya2008facilitation}%
  \BibitemOpen
  \bibfield  {author} {\bibinfo {author} {\bibfnamefont {S.~M.}\ \bibnamefont
  {Bhattacharyya}}, \bibinfo {author} {\bibfnamefont {B.}~\bibnamefont
  {Bagchi}}, \ and\ \bibinfo {author} {\bibfnamefont {P.~G.}\ \bibnamefont
  {Wolynes}},\ }\href {https://www.pnas.org/content/105/42/16077} {\bibfield
  {journal} {\bibinfo  {journal} {Proceedings of the National Academy of
  Sciences}\ }\textbf {\bibinfo {volume} {105}},\ \bibinfo {pages} {16077}
  (\bibinfo {year} {2008})}\BibitemShut {NoStop}%
\bibitem [{\citenamefont {Biroli}\ and\ \citenamefont
  {Bouchaud}(2012)}]{biroli2012random}%
  \BibitemOpen
  \bibfield  {author} {\bibinfo {author} {\bibfnamefont {G.}~\bibnamefont
  {Biroli}}\ and\ \bibinfo {author} {\bibfnamefont {J.-P.}\ \bibnamefont
  {Bouchaud}},\ }\href
  {https://onlinelibrary.wiley.com/doi/abs/10.1002/9781118202470.ch2}
  {\bibfield  {journal} {\bibinfo  {journal} {Structural Glasses and
  Supercooled Liquids: Theory, Experiment, and Applications}\ ,\ \bibinfo
  {pages} {31}} (\bibinfo {year} {2012})}\BibitemShut {NoStop}%
\bibitem [{\citenamefont {Berthier}\ \emph
  {et~al.}(2019{\natexlab{a}})\citenamefont {Berthier}, \citenamefont {Biroli},
  \citenamefont {Bouchaud},\ and\ \citenamefont {Tarjus}}]{berthier2019can}%
  \BibitemOpen
  \bibfield  {author} {\bibinfo {author} {\bibfnamefont {L.}~\bibnamefont
  {Berthier}}, \bibinfo {author} {\bibfnamefont {G.}~\bibnamefont {Biroli}},
  \bibinfo {author} {\bibfnamefont {J.-P.}\ \bibnamefont {Bouchaud}}, \ and\
  \bibinfo {author} {\bibfnamefont {G.}~\bibnamefont {Tarjus}},\ }\href
  {\doibase 10.1063/1.5086509} {\bibfield  {journal} {\bibinfo  {journal} {The
  Journal of Chemical Physics}\ }\textbf {\bibinfo {volume} {150}},\ \bibinfo
  {pages} {094501} (\bibinfo {year} {2019}{\natexlab{a}})}\BibitemShut
  {NoStop}%
\bibitem [{\citenamefont {Guiselin}\ \emph {et~al.}(2021)\citenamefont
  {Guiselin}, \citenamefont {Scalliet},\ and\ \citenamefont
  {Berthier}}]{shortwings}%
  \BibitemOpen
  \bibfield  {author} {\bibinfo {author} {\bibfnamefont {B.}~\bibnamefont
  {Guiselin}}, \bibinfo {author} {\bibfnamefont {C.}~\bibnamefont {Scalliet}},
  \ and\ \bibinfo {author} {\bibfnamefont {L.}~\bibnamefont {Berthier}},\
  }\href {https://arxiv.org/abs/2103.01569} {\bibfield  {journal} {\bibinfo
  {journal} {arXiv preprint arXiv:2103.01569}\ } (\bibinfo {year}
  {2021})}\BibitemShut {NoStop}%
\bibitem [{\citenamefont {Berthier}\ \emph {et~al.}(2016)\citenamefont
  {Berthier}, \citenamefont {Coslovich}, \citenamefont {Ninarello},\ and\
  \citenamefont {Ozawa}}]{berthier2016equilibrium}%
  \BibitemOpen
  \bibfield  {author} {\bibinfo {author} {\bibfnamefont {L.}~\bibnamefont
  {Berthier}}, \bibinfo {author} {\bibfnamefont {D.}~\bibnamefont {Coslovich}},
  \bibinfo {author} {\bibfnamefont {A.}~\bibnamefont {Ninarello}}, \ and\
  \bibinfo {author} {\bibfnamefont {M.}~\bibnamefont {Ozawa}},\ }\href
  {\doibase 10.1103/PhysRevLett.116.238002} {\bibfield  {journal} {\bibinfo
  {journal} {Phys. Rev. Lett.}\ }\textbf {\bibinfo {volume} {116}},\ \bibinfo
  {pages} {238002} (\bibinfo {year} {2016})}\BibitemShut {NoStop}%
\bibitem [{\citenamefont {Ninarello}\ \emph {et~al.}(2017)\citenamefont
  {Ninarello}, \citenamefont {Berthier},\ and\ \citenamefont
  {Coslovich}}]{swap}%
  \BibitemOpen
  \bibfield  {author} {\bibinfo {author} {\bibfnamefont {A.}~\bibnamefont
  {Ninarello}}, \bibinfo {author} {\bibfnamefont {L.}~\bibnamefont {Berthier}},
  \ and\ \bibinfo {author} {\bibfnamefont {D.}~\bibnamefont {Coslovich}},\
  }\href {\doibase 10.1103/PhysRevX.7.021039} {\bibfield  {journal} {\bibinfo
  {journal} {Phys. Rev. X}\ }\textbf {\bibinfo {volume} {7}},\ \bibinfo {pages}
  {021039} (\bibinfo {year} {2017})}\BibitemShut {NoStop}%
\bibitem [{\citenamefont {Berthier}\ \emph
  {et~al.}(2019{\natexlab{b}})\citenamefont {Berthier}, \citenamefont
  {Flenner}, \citenamefont {Fullerton}, \citenamefont {Scalliet},\ and\
  \citenamefont {Singh}}]{berthier2019efficient}%
  \BibitemOpen
  \bibfield  {author} {\bibinfo {author} {\bibfnamefont {L.}~\bibnamefont
  {Berthier}}, \bibinfo {author} {\bibfnamefont {E.}~\bibnamefont {Flenner}},
  \bibinfo {author} {\bibfnamefont {C.~J.}\ \bibnamefont {Fullerton}}, \bibinfo
  {author} {\bibfnamefont {C.}~\bibnamefont {Scalliet}}, \ and\ \bibinfo
  {author} {\bibfnamefont {M.}~\bibnamefont {Singh}},\ }\href {\doibase
  10.1088/1742-5468/ab1910} {\bibfield  {journal} {\bibinfo  {journal} {Journal
  of Statistical Mechanics: Theory and Experiment}\ }\textbf {\bibinfo {volume}
  {2019}},\ \bibinfo {pages} {064004} (\bibinfo {year}
  {2019}{\natexlab{b}})}\BibitemShut {NoStop}%
\bibitem [{\citenamefont {Berthier}\ \emph {et~al.}(2011)\citenamefont
  {Berthier}, \citenamefont {Biroli}, \citenamefont {Bouchaud}, \citenamefont
  {Cipelletti},\ and\ \citenamefont {van Saarloos}}]{berthier2011dynamical}%
  \BibitemOpen
  \bibfield  {author} {\bibinfo {author} {\bibfnamefont {L.}~\bibnamefont
  {Berthier}}, \bibinfo {author} {\bibfnamefont {G.}~\bibnamefont {Biroli}},
  \bibinfo {author} {\bibfnamefont {J.-P.}\ \bibnamefont {Bouchaud}}, \bibinfo
  {author} {\bibfnamefont {L.}~\bibnamefont {Cipelletti}}, \ and\ \bibinfo
  {author} {\bibfnamefont {W.}~\bibnamefont {van Saarloos}},\ }\href@noop {}
  {\emph {\bibinfo {title} {Dynamical heterogeneities in glasses, colloids, and
  granular media}}},\ Vol.\ \bibinfo {volume} {150}\ (\bibinfo  {publisher}
  {OUP Oxford},\ \bibinfo {year} {2011})\BibitemShut {NoStop}%
\bibitem [{\citenamefont {Berthier}(2011)}]{berthier2011physics}%
  \BibitemOpen
  \bibfield  {author} {\bibinfo {author} {\bibfnamefont {L.}~\bibnamefont
  {Berthier}},\ }\href {https://physics.aps.org/articles/v4/42} {\bibfield
  {journal} {\bibinfo  {journal} {Physics}\ }\textbf {\bibinfo {volume} {4}},\
  \bibinfo {pages} {42} (\bibinfo {year} {2011})}\BibitemShut {NoStop}%
\bibitem [{\citenamefont {Brawer}(1984)}]{brawer1984theory}%
  \BibitemOpen
  \bibfield  {author} {\bibinfo {author} {\bibfnamefont {S.}~\bibnamefont
  {Brawer}},\ }\href {https://doi.org/10.1063/1.447697} {\bibfield  {journal}
  {\bibinfo  {journal} {The Journal of chemical physics}\ }\textbf {\bibinfo
  {volume} {81}},\ \bibinfo {pages} {954} (\bibinfo {year} {1984})}\BibitemShut
  {NoStop}%
\bibitem [{\citenamefont {Dyre}(1987)}]{dyre1987master}%
  \BibitemOpen
  \bibfield  {author} {\bibinfo {author} {\bibfnamefont {J.~C.}\ \bibnamefont
  {Dyre}},\ }\href {\doibase 10.1103/PhysRevLett.58.792} {\bibfield  {journal}
  {\bibinfo  {journal} {Phys. Rev. Lett.}\ }\textbf {\bibinfo {volume} {58}},\
  \bibinfo {pages} {792} (\bibinfo {year} {1987})}\BibitemShut {NoStop}%
\bibitem [{\citenamefont {B\"assler}(1987)}]{bassler1987viscous}%
  \BibitemOpen
  \bibfield  {author} {\bibinfo {author} {\bibfnamefont {H.}~\bibnamefont
  {B\"assler}},\ }\href {\doibase 10.1103/PhysRevLett.58.767} {\bibfield
  {journal} {\bibinfo  {journal} {Phys. Rev. Lett.}\ }\textbf {\bibinfo
  {volume} {58}},\ \bibinfo {pages} {767} (\bibinfo {year} {1987})}\BibitemShut
  {NoStop}%
\bibitem [{\citenamefont {Bouchaud}(1992)}]{bouchaud1992weak}%
  \BibitemOpen
  \bibfield  {author} {\bibinfo {author} {\bibfnamefont {J.-P.}\ \bibnamefont
  {Bouchaud}},\ }\href {https://doi.org/10.1051/jp1:1992238} {\bibfield
  {journal} {\bibinfo  {journal} {Journal de Physique I}\ }\textbf {\bibinfo
  {volume} {2}},\ \bibinfo {pages} {1705} (\bibinfo {year} {1992})}\BibitemShut
  {NoStop}%
\bibitem [{\citenamefont {Monthus}\ and\ \citenamefont
  {Bouchaud}(1996)}]{Monthus_1996}%
  \BibitemOpen
  \bibfield  {author} {\bibinfo {author} {\bibfnamefont {C.}~\bibnamefont
  {Monthus}}\ and\ \bibinfo {author} {\bibfnamefont {J.-P.}\ \bibnamefont
  {Bouchaud}},\ }\href {\doibase 10.1088/0305-4470/29/14/012} {\bibfield
  {journal} {\bibinfo  {journal} {Journal of Physics A: Mathematical and
  General}\ }\textbf {\bibinfo {volume} {29}},\ \bibinfo {pages} {3847}
  (\bibinfo {year} {1996})}\BibitemShut {NoStop}%
\bibitem [{\citenamefont {{Jean-Philippe Bouchaud}}\ \emph
  {et~al.}(1995)\citenamefont {{Jean-Philippe Bouchaud}}, \citenamefont {{Alain
  Comtet}},\ and\ \citenamefont {{C\'ecile Monthus}}}]{comtetmonthus}%
  \BibitemOpen
  \bibfield  {author} {\bibinfo {author} {\bibnamefont {{Jean-Philippe
  Bouchaud}}}, \bibinfo {author} {\bibnamefont {{Alain Comtet}}}, \ and\
  \bibinfo {author} {\bibnamefont {{C\'ecile Monthus}}},\ }\href {\doibase
  10.1051/jp1:1995104} {\bibfield  {journal} {\bibinfo  {journal} {J. Phys. I
  France}\ }\textbf {\bibinfo {volume} {5}},\ \bibinfo {pages} {1521} (\bibinfo
  {year} {1995})}\BibitemShut {NoStop}%
\bibitem [{\citenamefont {Denny}\ \emph {et~al.}(2003)\citenamefont {Denny},
  \citenamefont {Reichman},\ and\ \citenamefont {Bouchaud}}]{denny2003trap}%
  \BibitemOpen
  \bibfield  {author} {\bibinfo {author} {\bibfnamefont {R.~A.}\ \bibnamefont
  {Denny}}, \bibinfo {author} {\bibfnamefont {D.~R.}\ \bibnamefont {Reichman}},
  \ and\ \bibinfo {author} {\bibfnamefont {J.-P.}\ \bibnamefont {Bouchaud}},\
  }\href {\doibase 10.1103/PhysRevLett.90.025503} {\bibfield  {journal}
  {\bibinfo  {journal} {Phys. Rev. Lett.}\ }\textbf {\bibinfo {volume} {90}},\
  \bibinfo {pages} {025503} (\bibinfo {year} {2003})}\BibitemShut {NoStop}%
\bibitem [{\citenamefont {Heuer}\ \emph {et~al.}(2005)\citenamefont {Heuer},
  \citenamefont {Doliwa},\ and\ \citenamefont
  {Saksaengwijit}}]{heuer2005potential}%
  \BibitemOpen
  \bibfield  {author} {\bibinfo {author} {\bibfnamefont {A.}~\bibnamefont
  {Heuer}}, \bibinfo {author} {\bibfnamefont {B.}~\bibnamefont {Doliwa}}, \
  and\ \bibinfo {author} {\bibfnamefont {A.}~\bibnamefont {Saksaengwijit}},\
  }\href {\doibase 10.1103/PhysRevE.72.021503} {\bibfield  {journal} {\bibinfo
  {journal} {Phys. Rev. E}\ }\textbf {\bibinfo {volume} {72}},\ \bibinfo
  {pages} {021503} (\bibinfo {year} {2005})}\BibitemShut {NoStop}%
\bibitem [{\citenamefont {Diezemann}\ and\ \citenamefont
  {Heuer}(2011)}]{diezemann2011memory}%
  \BibitemOpen
  \bibfield  {author} {\bibinfo {author} {\bibfnamefont {G.}~\bibnamefont
  {Diezemann}}\ and\ \bibinfo {author} {\bibfnamefont {A.}~\bibnamefont
  {Heuer}},\ }\href {\doibase 10.1103/PhysRevE.83.031505} {\bibfield  {journal}
  {\bibinfo  {journal} {Phys. Rev. E}\ }\textbf {\bibinfo {volume} {83}},\
  \bibinfo {pages} {031505} (\bibinfo {year} {2011})}\BibitemShut {NoStop}%
\bibitem [{\citenamefont {Diezemann}(1997)}]{diezemann1997free}%
  \BibitemOpen
  \bibfield  {author} {\bibinfo {author} {\bibfnamefont {G.}~\bibnamefont
  {Diezemann}},\ }\href {\doibase 10.1063/1.474148} {\bibfield  {journal}
  {\bibinfo  {journal} {The Journal of Chemical Physics}\ }\textbf {\bibinfo
  {volume} {107}},\ \bibinfo {pages} {10112} (\bibinfo {year}
  {1997})}\BibitemShut {NoStop}%
\bibitem [{\citenamefont {Rehwald}\ \emph {et~al.}(2010)\citenamefont
  {Rehwald}, \citenamefont {Rubner},\ and\ \citenamefont
  {Heuer}}]{rehwald2010coupled}%
  \BibitemOpen
  \bibfield  {author} {\bibinfo {author} {\bibfnamefont {C.}~\bibnamefont
  {Rehwald}}, \bibinfo {author} {\bibfnamefont {O.}~\bibnamefont {Rubner}}, \
  and\ \bibinfo {author} {\bibfnamefont {A.}~\bibnamefont {Heuer}},\ }\href
  {\doibase 10.1103/PhysRevLett.105.117801} {\bibfield  {journal} {\bibinfo
  {journal} {Phys. Rev. Lett.}\ }\textbf {\bibinfo {volume} {105}},\ \bibinfo
  {pages} {117801} (\bibinfo {year} {2010})}\BibitemShut {NoStop}%
\bibitem [{\citenamefont {Arkhipov}\ and\ \citenamefont
  {Baessler}(1994)}]{arkhipov1994random}%
  \BibitemOpen
  \bibfield  {author} {\bibinfo {author} {\bibfnamefont {V.~I.}\ \bibnamefont
  {Arkhipov}}\ and\ \bibinfo {author} {\bibfnamefont {H.}~\bibnamefont
  {Baessler}},\ }\href {\doibase 10.1021/j100053a047} {\bibfield  {journal}
  {\bibinfo  {journal} {The Journal of Physical Chemistry}\ }\textbf {\bibinfo
  {volume} {98}},\ \bibinfo {pages} {662} (\bibinfo {year} {1994})}\BibitemShut
  {NoStop}%
\bibitem [{\citenamefont {Arkhipov}\ \emph {et~al.}(1996)\citenamefont
  {Arkhipov}, \citenamefont {B{\"a}ssler},\ and\ \citenamefont
  {Khramtchenkov}}]{arkhipov1996random}%
  \BibitemOpen
  \bibfield  {author} {\bibinfo {author} {\bibfnamefont {V.}~\bibnamefont
  {Arkhipov}}, \bibinfo {author} {\bibfnamefont {H.}~\bibnamefont
  {B{\"a}ssler}}, \ and\ \bibinfo {author} {\bibfnamefont {D.}~\bibnamefont
  {Khramtchenkov}},\ }\href {https://doi.org/10.1021/jp9522831} {\bibfield
  {journal} {\bibinfo  {journal} {The Journal of Physical Chemistry}\ }\textbf
  {\bibinfo {volume} {100}},\ \bibinfo {pages} {5118} (\bibinfo {year}
  {1996})}\BibitemShut {NoStop}%
\bibitem [{\citenamefont {Bergroth}\ \emph {et~al.}(2005)\citenamefont
  {Bergroth}, \citenamefont {Vogel},\ and\ \citenamefont
  {Glotzer}}]{bergroth2005examination}%
  \BibitemOpen
  \bibfield  {author} {\bibinfo {author} {\bibfnamefont {M.~N.}\ \bibnamefont
  {Bergroth}}, \bibinfo {author} {\bibfnamefont {M.}~\bibnamefont {Vogel}}, \
  and\ \bibinfo {author} {\bibfnamefont {S.~C.}\ \bibnamefont {Glotzer}},\
  }\href {https://doi.org/10.1021/jp0447946} {\bibfield  {journal} {\bibinfo
  {journal} {The Journal of Physical Chemistry B}\ }\textbf {\bibinfo {volume}
  {109}},\ \bibinfo {pages} {6748} (\bibinfo {year} {2005})}\BibitemShut
  {NoStop}%
\bibitem [{\citenamefont {Vogel}\ and\ \citenamefont
  {Glotzer}(2004)}]{vogel2004spatially}%
  \BibitemOpen
  \bibfield  {author} {\bibinfo {author} {\bibfnamefont {M.}~\bibnamefont
  {Vogel}}\ and\ \bibinfo {author} {\bibfnamefont {S.~C.}\ \bibnamefont
  {Glotzer}},\ }\href {\doibase 10.1103/PhysRevLett.92.255901} {\bibfield
  {journal} {\bibinfo  {journal} {Phys. Rev. Lett.}\ }\textbf {\bibinfo
  {volume} {92}},\ \bibinfo {pages} {255901} (\bibinfo {year}
  {2004})}\BibitemShut {NoStop}%
\bibitem [{\citenamefont {Candelier}\ \emph {et~al.}(2009)\citenamefont
  {Candelier}, \citenamefont {Dauchot},\ and\ \citenamefont
  {Biroli}}]{candelier2009building}%
  \BibitemOpen
  \bibfield  {author} {\bibinfo {author} {\bibfnamefont {R.}~\bibnamefont
  {Candelier}}, \bibinfo {author} {\bibfnamefont {O.}~\bibnamefont {Dauchot}},
  \ and\ \bibinfo {author} {\bibfnamefont {G.}~\bibnamefont {Biroli}},\ }\href
  {\doibase 10.1103/PhysRevLett.102.088001} {\bibfield  {journal} {\bibinfo
  {journal} {Phys. Rev. Lett.}\ }\textbf {\bibinfo {volume} {102}},\ \bibinfo
  {pages} {088001} (\bibinfo {year} {2009})}\BibitemShut {NoStop}%
\bibitem [{\citenamefont {Candelier}\ \emph {et~al.}(2010)\citenamefont
  {Candelier}, \citenamefont {Widmer-Cooper}, \citenamefont {Kummerfeld},
  \citenamefont {Dauchot}, \citenamefont {Biroli}, \citenamefont {Harrowell},\
  and\ \citenamefont {Reichman}}]{candelier2010spatiotemporal}%
  \BibitemOpen
  \bibfield  {author} {\bibinfo {author} {\bibfnamefont {R.}~\bibnamefont
  {Candelier}}, \bibinfo {author} {\bibfnamefont {A.}~\bibnamefont
  {Widmer-Cooper}}, \bibinfo {author} {\bibfnamefont {J.~K.}\ \bibnamefont
  {Kummerfeld}}, \bibinfo {author} {\bibfnamefont {O.}~\bibnamefont {Dauchot}},
  \bibinfo {author} {\bibfnamefont {G.}~\bibnamefont {Biroli}}, \bibinfo
  {author} {\bibfnamefont {P.}~\bibnamefont {Harrowell}}, \ and\ \bibinfo
  {author} {\bibfnamefont {D.~R.}\ \bibnamefont {Reichman}},\ }\href {\doibase
  10.1103/PhysRevLett.105.135702} {\bibfield  {journal} {\bibinfo  {journal}
  {Phys. Rev. Lett.}\ }\textbf {\bibinfo {volume} {105}},\ \bibinfo {pages}
  {135702} (\bibinfo {year} {2010})}\BibitemShut {NoStop}%
\bibitem [{\citenamefont {Keys}\ \emph {et~al.}(2011)\citenamefont {Keys},
  \citenamefont {Hedges}, \citenamefont {Garrahan}, \citenamefont {Glotzer},\
  and\ \citenamefont {Chandler}}]{keys2011excitations}%
  \BibitemOpen
  \bibfield  {author} {\bibinfo {author} {\bibfnamefont {A.~S.}\ \bibnamefont
  {Keys}}, \bibinfo {author} {\bibfnamefont {L.~O.}\ \bibnamefont {Hedges}},
  \bibinfo {author} {\bibfnamefont {J.~P.}\ \bibnamefont {Garrahan}}, \bibinfo
  {author} {\bibfnamefont {S.~C.}\ \bibnamefont {Glotzer}}, \ and\ \bibinfo
  {author} {\bibfnamefont {D.}~\bibnamefont {Chandler}},\ }\href {\doibase
  10.1103/PhysRevX.1.021013} {\bibfield  {journal} {\bibinfo  {journal} {Phys.
  Rev. X}\ }\textbf {\bibinfo {volume} {1}},\ \bibinfo {pages} {021013}
  (\bibinfo {year} {2011})}\BibitemShut {NoStop}%
\bibitem [{\citenamefont {Schoenholz}\ \emph {et~al.}(2016)\citenamefont
  {Schoenholz}, \citenamefont {Cubuk}, \citenamefont {Sussman}, \citenamefont
  {Kaxiras},\ and\ \citenamefont {Liu}}]{schoenholz2016structural}%
  \BibitemOpen
  \bibfield  {author} {\bibinfo {author} {\bibfnamefont {S.~S.}\ \bibnamefont
  {Schoenholz}}, \bibinfo {author} {\bibfnamefont {E.~D.}\ \bibnamefont
  {Cubuk}}, \bibinfo {author} {\bibfnamefont {D.~M.}\ \bibnamefont {Sussman}},
  \bibinfo {author} {\bibfnamefont {E.}~\bibnamefont {Kaxiras}}, \ and\
  \bibinfo {author} {\bibfnamefont {A.~J.}\ \bibnamefont {Liu}},\ }\href
  {https://www.nature.com/articles/nphys3644} {\bibfield  {journal} {\bibinfo
  {journal} {Nature Physics}\ }\textbf {\bibinfo {volume} {12}},\ \bibinfo
  {pages} {469} (\bibinfo {year} {2016})}\BibitemShut {NoStop}%
\bibitem [{\citenamefont {Chacko}\ \emph {et~al.}(2021)\citenamefont {Chacko},
  \citenamefont {Landes}, \citenamefont {Biroli}, \citenamefont {Dauchot},
  \citenamefont {Liu},\ and\ \citenamefont {Reichman}}]{chacko}%
  \BibitemOpen
  \bibfield  {author} {\bibinfo {author} {\bibfnamefont {R.~N.}\ \bibnamefont
  {Chacko}}, \bibinfo {author} {\bibfnamefont {F.~P.}\ \bibnamefont {Landes}},
  \bibinfo {author} {\bibfnamefont {G.}~\bibnamefont {Biroli}}, \bibinfo
  {author} {\bibfnamefont {O.}~\bibnamefont {Dauchot}}, \bibinfo {author}
  {\bibfnamefont {A.~J.}\ \bibnamefont {Liu}}, \ and\ \bibinfo {author}
  {\bibfnamefont {D.~R.}\ \bibnamefont {Reichman}},\ }\href
  {https://arxiv.org/abs/2103.01852} {\bibfield  {journal} {\bibinfo  {journal}
  {arXiv preprint arXiv:2103.01852}\ } (\bibinfo {year} {2021})}\BibitemShut
  {NoStop}%
\bibitem [{\citenamefont {Diezemann}(2005)}]{diezemann2005aging}%
  \BibitemOpen
  \bibfield  {author} {\bibinfo {author} {\bibfnamefont {G.}~\bibnamefont
  {Diezemann}},\ }\href {https://doi.org/10.1063/1.2128700} {\bibfield
  {journal} {\bibinfo  {journal} {The Journal of chemical physics}\ }\textbf
  {\bibinfo {volume} {123}},\ \bibinfo {pages} {204510} (\bibinfo {year}
  {2005})}\BibitemShut {NoStop}%
\bibitem [{\citenamefont {Rehwald}\ and\ \citenamefont
  {Heuer}(2012)}]{rehwald2012coupled}%
  \BibitemOpen
  \bibfield  {author} {\bibinfo {author} {\bibfnamefont {C.}~\bibnamefont
  {Rehwald}}\ and\ \bibinfo {author} {\bibfnamefont {A.}~\bibnamefont
  {Heuer}},\ }\href {\doibase 10.1103/PhysRevE.86.051504} {\bibfield  {journal}
  {\bibinfo  {journal} {Phys. Rev. E}\ }\textbf {\bibinfo {volume} {86}},\
  \bibinfo {pages} {051504} (\bibinfo {year} {2012})}\BibitemShut {NoStop}%
\bibitem [{\citenamefont {B{\"o}hmer}\ \emph {et~al.}(1998)\citenamefont
  {B{\"o}hmer}, \citenamefont {Chamberlin}, \citenamefont {Diezemann},
  \citenamefont {Geil}, \citenamefont {Heuer}, \citenamefont {Hinze},
  \citenamefont {Kuebler}, \citenamefont {Richert}, \citenamefont {Schiener},
  \citenamefont {Sillescu} \emph {et~al.}}]{bohmer1998nature}%
  \BibitemOpen
  \bibfield  {author} {\bibinfo {author} {\bibfnamefont {R.}~\bibnamefont
  {B{\"o}hmer}}, \bibinfo {author} {\bibfnamefont {R.}~\bibnamefont
  {Chamberlin}}, \bibinfo {author} {\bibfnamefont {G.}~\bibnamefont
  {Diezemann}}, \bibinfo {author} {\bibfnamefont {B.}~\bibnamefont {Geil}},
  \bibinfo {author} {\bibfnamefont {A.}~\bibnamefont {Heuer}}, \bibinfo
  {author} {\bibfnamefont {G.}~\bibnamefont {Hinze}}, \bibinfo {author}
  {\bibfnamefont {S.}~\bibnamefont {Kuebler}}, \bibinfo {author} {\bibfnamefont
  {R.}~\bibnamefont {Richert}}, \bibinfo {author} {\bibfnamefont
  {B.}~\bibnamefont {Schiener}}, \bibinfo {author} {\bibfnamefont
  {H.}~\bibnamefont {Sillescu}},  \emph {et~al.},\ }\href {\doibase
  https://doi.org/10.1016/S0022-3093(98)00581-X} {\bibfield  {journal}
  {\bibinfo  {journal} {Journal of non-crystalline solids}\ }\textbf {\bibinfo
  {volume} {235}},\ \bibinfo {pages} {1} (\bibinfo {year} {1998})}\BibitemShut
  {NoStop}%
\bibitem [{\citenamefont {Marsaglia}\ and\ \citenamefont
  {Tsang}(2000)}]{gamma}%
  \BibitemOpen
  \bibfield  {author} {\bibinfo {author} {\bibfnamefont {G.}~\bibnamefont
  {Marsaglia}}\ and\ \bibinfo {author} {\bibfnamefont {W.~W.}\ \bibnamefont
  {Tsang}},\ }\href {\doibase 10.1145/358407.358414} {\bibfield  {journal}
  {\bibinfo  {journal} {ACM Trans. Math. Softw.}\ }\textbf {\bibinfo {volume}
  {26}},\ \bibinfo {pages} {363} (\bibinfo {year} {2000})}\BibitemShut
  {NoStop}%
\bibitem [{\citenamefont {Gardiner}\ \emph {et~al.}(1985)\citenamefont
  {Gardiner} \emph {et~al.}}]{gardiner1985handbook}%
  \BibitemOpen
  \bibfield  {author} {\bibinfo {author} {\bibfnamefont {C.~W.}\ \bibnamefont
  {Gardiner}} \emph {et~al.},\ }\href@noop {} {\emph {\bibinfo {title}
  {Handbook of stochastic methods}}},\ Vol.~\bibinfo {volume} {3}\ (\bibinfo
  {publisher} {springer Berlin},\ \bibinfo {year} {1985})\BibitemShut {NoStop}%
\bibitem [{\citenamefont {Berthier}(2020)}]{berthier2020self}%
  \BibitemOpen
  \bibfield  {author} {\bibinfo {author} {\bibfnamefont {L.}~\bibnamefont
  {Berthier}},\ }\href {https://arxiv.org/abs/2010.12244} {\bibfield  {journal}
  {\bibinfo  {journal} {arXiv preprint arXiv:2010.12244}\ } (\bibinfo {year}
  {2020})}\BibitemShut {NoStop}%
\bibitem [{\citenamefont {Schneider}\ \emph {et~al.}(2000)\citenamefont
  {Schneider}, \citenamefont {Brand}, \citenamefont {Lunkenheimer},\ and\
  \citenamefont {Loidl}}]{schneider2000excess}%
  \BibitemOpen
  \bibfield  {author} {\bibinfo {author} {\bibfnamefont {U.}~\bibnamefont
  {Schneider}}, \bibinfo {author} {\bibfnamefont {R.}~\bibnamefont {Brand}},
  \bibinfo {author} {\bibfnamefont {P.}~\bibnamefont {Lunkenheimer}}, \ and\
  \bibinfo {author} {\bibfnamefont {A.}~\bibnamefont {Loidl}},\ }\href
  {\doibase 10.1103/PhysRevLett.84.5560} {\bibfield  {journal} {\bibinfo
  {journal} {Phys. Rev. Lett.}\ }\textbf {\bibinfo {volume} {84}},\ \bibinfo
  {pages} {5560} (\bibinfo {year} {2000})}\BibitemShut {NoStop}%
\end{thebibliography}%

\end{document}